\documentclass[prd,groupedaddress, showpacs, doublecolumn,nofootinbib]{revtex4} %
\usepackage{graphicx}
\usepackage{dcolumn}
\usepackage{bm}
\usepackage{amssymb}
\usepackage{amsmath}
\usepackage{epsfig}
\usepackage{hhline}
\usepackage{color}
\usepackage{multirow}
\usepackage{cancel}

\def\lapp{\mathrel{\rlap{\raise.5ex\hbox{$<$}}
                    {\lower.5ex\hbox{$\sim$}}}}
\def\gapp{\mathrel{\rlap{\raise.5ex\hbox{$>$}}
                    {\lower.5ex\hbox{$\sim$}}}}

{
{

\newcommand{\bmt}{\begin{pmatrix}}
\newcommand{\emt}{\end{pmatrix}}
\newcommand{\ba}{\begin{array}{c}}
\newcommand{\ea}{\end{array}}
\newcommand{\be}{\begin{equation}}
\newcommand{\ee}{\end{equation}}
\newcommand{\bea}{\begin{eqnarray}}
\newcommand{\eea}{\end{eqnarray}}

\newcommand{\bi}{\begin{itemize}}
\newcommand{\ei}{\end{itemize}}

\newcommand{\baz}{\begin{array}{cc}}

\newcommand{\mathsym}[1]{{}}

\newcommand{\bt}{\begin{tabular}}
\newcommand{\et}{\end{tabular}}

\newcommand{\benu}{\begin{enumerate}}
\newcommand{\eenu}{\end{enumerate}}
\newcommand{\bav}{\begin{array}{cccc}}

\begin{document}
\title{\bf Singlet-Doublet Fermionic Dark Matter, Neutrino Mass and Collider Signatures}

\author{Subhaditya Bhattacharya}
\email{subhab@iitg.ernet.in}
\affiliation{Department of Physics, Indian Institute of Technology Guwahati, North Guwahati, Assam- 781039, India}
\author{Nirakar Sahoo}
\email{nirakar.pintu.sahoo@gmail.com}
\author{Narendra Sahu}
\email{nsahu@iith.ac.in}
\affiliation{Department of Physics, Indian Institute of Technology,
  Hyderabad, Kandi, Sangareddy, 502285, 
Telangana, India}

\begin{abstract}
We propose a minimal extension of the standard model (SM) by including a scalar triplet with hypercharge 
2 and two vector-like leptons: one doublet and a singlet, to explain simultaneously the non-zero neutrino 
mass and dark matter (DM) content of the Universe. The DM emerges out as a mixture of the neutral component 
of vector-like lepton doublet and singlet, being odd under a discrete $Z_2$ symmetry. After electroweak 
symmetry breaking the triplet scalar gets an induced vev, which give Majorana masses not only to the light 
neutrinos but also to the DM. Due to the Majorana mass of DM, the $Z$ mediated elastic scattering with 
nucleon is forbidden. However, the Higgs mediated direct detection cross-section of the DM gives an excellent 
opportunity to probe it at Xenon-1T. The DM can not be detected at collider. However, the charged partner of the 
DM (often next-to-lightest stable particle) can give large displaced vertex signature at the Large Hadron 
Collider (LHC).

\end{abstract}

\pacs{98.80.Cq , 12.60.Fr}
\maketitle

\newpage

\section{Introduction}
Astrophysical evidences like galaxy rotation curves, gravitational lensing, large scale structure of the universe and 
anisotropies in Cosmic Microwave Background Radiation (CMBR) hint towards the existence of an unknown form of
non-luminous matter, called dark matter (DM) in our Universe~\cite{DM_review1,DM_review2}. However they imply only about 
the gravitational property of DM, whose relic abundance is precisely measured by the satellite borne experiments such as 
WMAP~\cite{wmap} and PLANCK~\cite{PLANCK} to be $\Omega_{\rm DM}h^2 = 0.1199\pm 0.0027$. However, the microscopic nature of DM is 
still a mystery. In fact, none of the particles in the standard model (SM) mimics the properties of DM. This leads to a rich 
possibility of DM model building in the beyond SM physics, though the WIMP (Weakly Interacting Massive Particle) paradigm is 
the most popular one. 

Another significant hint of physics beyond the SM came through the discovery of non-zero masses of
left-handed neutrinos in the last decade. Long baseline neutrino oscillation experiments~\cite{Fukuda:2001nk, Bilenky:1980cx} confirmed that neutrinos
have sub-eV masses and they mix among themselves. This also has compelled particle physicists to explore structures
beyond the SM. There are two major issues involving neutrinos: (i) their nature, Dirac or Majorana, (ii) mass
hierarchy, normal or inverted. A popular solution for non-zero Majorana masses of active neutrinos is to introduce the 
seesaw mechanisms~\cite{type_I,type_II,type_III} which are different realisations of the dimension five operator~\cite{dim-5-operator}: 
${LLHH}/{\Lambda}$, where $L$ and $H$ are the lepton and Higgs doublet of the SM and $\Lambda$ is the scale of new physics. After 
electroweak (EW) symmetry breaking the neutrino mass is given by $M_\nu =\langle H \rangle^2/\Lambda$. Thus for tiny neutrino mass 
$M_\nu \sim 0.1$ eV, the new physics scale requires to be pretty heavy $\Lambda \sim 10^{14}$ GeV when the involved couplings are 
order 1. However, $\Lambda$ can be reduced down to TeV scales if the couplings are assumed to be smaller.

In an attempt to bring dark matter and neutrino mass mechanisms under one umbrella (for some earlier attempts, see~\cite{Bhattacharya:2016rqj, Bhattacharya:2016lts, Bhattacharya:2016qsg,Sahu:2007uh,McDonald:2007ka,Sahu:2008aw,Chatterjee:2014vua,Patra:2014sua,Patra:2016shz} ),  we consider a minimal type-II seesaw extension of the SM by adding a TeV scale triplet scalar $\Delta$ with hypercharge 2 and introduce two additional vector-like leptons: one doublet $N \equiv (N^0, N^-)^T$ 
and a singlet $\chi$. A $Z_2$ symmetry is also imposed under which $N$ and $\chi$ are odd while all other fields are even. As a result the DM 
emerges out to be a mixed state of singlet and neutral component of the doublet vector-like leptons. Such DM frameworks have been discussed earlier; 
see for example, refs.~\cite{ArkaniHamed:2005yv,Mahbubani:2005pt,D'Eramo:2007ga,Enberg:2007rp,Cynolter:2008ea,
Cohen:2011ec,Cheung:2013dua,Restrepo:2015ura,Calibbi:2015nha,Cynolter:2015sua,Bhattacharya:2015qpa, Bhattacharya:2016lyg}. However, the presence of the 
triplet adds to some interesting DM phenomenology as we will discuss in this paper. Since the scalar triplet can be light, it contributes to the 
relic abundance of DM through s-channel resonance on top of $Z$ and $H$ mediation. Moreover, it relaxes the strong constraints coming from direct 
detection. 

The triplet scalar not only couples to the SM lepton and Higgs doublets, but also to the additional vector-like lepton 
doublet $N$. The Majorana couplings of $\Delta$ with $N$, $L$ and $H$ is then be given by $f_N \Delta N N + f_L \Delta L L 
+ \mu \Delta^\dagger H H$. Note that if the triplet is heavier than the DM and leptons, it can be integrated out and hence 
effectively generating the dimension five operators: 
\begin{equation*}
 \left( \frac{LLHH}{\Lambda} + \frac{NNHH}{\Lambda} \right)\,,
\end{equation*}
where $\Lambda \sim M_\Delta$. After EW symmetry breaking $\Delta$ acquires an induced vacuum expectation value (vev) of ${\cal O} (1)$ 
GeV which in turn give Majorana masses to light neutrinos as well as to $N^0$. Since $N^0$ is a vector-like Dirac fermion, it can 
have a Dirac mass too. As a result $N^0$ splits up into two pseudo-Dirac fermions, with a mass splitting of sub-GeV order, whose 
elastic scattering with the nucleon mediated by $Z$-boson is forbidden. This feature of the model leads to a survival of larger 
region of parameter space from direct search constraints given by the latest data from Xenon-100~\cite{xenon100} and 
LUX~\cite{Akerib:2016vxi}. On the other hand, the Higgs mediated elastic scattering of the DM with the nucleon gives an 
excellent opportunity to detect it at future direct search experiments such as XENON1T~\cite{Aprile:2015uzo}. It is harder to see the 
signature of only DM production at collider as they need to recoil against an ISR jet for missing energy. However, the charged 
partner of the DM (which is next-to-lightest stable particle) can be produced copiously which eventually decays to DM giving rise 
to leptons and missing energy. More interestingly, the charged companion can also give large displaced vertex signature as we will elaborate. 

The paper is arranged as follows. In section \ref{model}, we discuss about model and formalism, mixing in fermionic and in 
scalar sector. In section \ref{neutrino}, we explain non-zero neutrino mass in a type II see-saw scenario while section IV 
is devoted to illustrate the pseudo-Dirac nature of DM. The relic abundance of DM is obtained in section \ref{relic}. The 
inelastic scattering of DM with the nucleus for direct search is presented in section \ref{DDin}. Section \ref{dd} is devoted for 
direct detection of DM through elastic scattering and limits on model parameter space. The displaced vertex signature of the 
charged partner of DM is discussed in sec \ref{dec}. With a summary of the analysis, we finally conclude in section \ref{con}.
\\

\section{The Model}\label{model}
As already been stated in the introduction, we extend the standard model (SM) by introducing two vector like fermions  
$N^T=(N^0, N^-)$ (1,2,-1) and $\chi^0$ (1,1,0) and a scalar triplet $\Delta$ (1,3,2), where the numbers inside the parenthesis are 
quantum numbers under the SM gauge group $SU(3)_c\times SU(2)_L\times U(1)_Y$. A $Z_2$ symmetry is imposed under which $\chi^0$ 
and $N$ are odd, while other fields are even. The relevant Lagrangian involving the additional fields is given by:
\begin{equation}\label{Lagrangian}
\mathcal{L}_{\rm new}=\overline{N} \cancel{D} N + \overline{\chi^0}
\cancel{\partial} \chi^0 + (D^\mu \Delta)^\dagger (D_\mu
\Delta) + M_N \overline{N}N + M_\chi \overline{\chi^0}\chi^0 +
\mathcal{L}_{yuk} - V(\Delta,H)\,,
\end{equation}
where $D_\mu$ is the covariant derivative involving $SU(2)$ ($W_\mu$) and $U(1)_Y (B_\mu)$ gauge bosons and is given by :
\begin{equation*}
D_\mu= \partial_\mu - i\frac{g}{2} \tau.W_\mu - ig^\prime \frac{Y}{2}B_\mu\,.
\end{equation*}
The scalar potential involving SM doublet ($H$) and triplet ($\Delta$) in Eq. (\ref{Lagrangian}) is given by 
 \begin{eqnarray}\label{eq:ScalarPotential}
\hspace*{-0.5cm}
 V(\Delta, H) &=&  -   \mu_H^2 H^\dagger H + \lambda_H (H^\dagger H)^2 +  \mu_{\Delta}^2 (\Delta^\dagger \Delta)+
 \lambda_\Delta (\Delta^\dagger \Delta)^2\nonumber\\
& +& \lambda_{H\Delta} (H^\dagger H) (\Delta^\dagger \Delta)+ 
\frac{1}{2}\left[\mu \Delta^\dagger H H + {\rm h.c.} \right]\,,
\end{eqnarray}
\\ where $\Delta$ in matrix form is
\begin{equation}
 \Delta = \begin{pmatrix}
           \frac{\Delta ^+}{\sqrt{2}} &  \Delta ^{+ +} \\
           \Delta^0 		  &  - \frac{\Delta^+}{\sqrt{2}}
          \end{pmatrix}\,.
\end{equation}
We assume that $\mu_\Delta ^2$ is positive. So it doesn't acquire a vacuum expectation value (vev). But it gets an induced vev 
after EW phase transition. The vev of $\Delta$ is given by
 \begin{equation}
  \langle \Delta \rangle \equiv u_\Delta  \approx - \frac{\mu
    v^2}{\sqrt{2}(\mu_\Delta^2 + \lambda_{H\Delta} v^2/2) } \label{vev}
 \end{equation}
where $v$ is the vev of Higgs field and its value is 174 GeV.

The Yukawa interaction in Eq. (\ref{Lagrangian}) is given by:
\begin{equation}\label{yukawa_coupling}
\mathcal{L}_{yuk}=\frac{1}{\sqrt{2}} \left[ (f_L)_{\alpha \beta} \overline{L_\alpha^c} i \tau_2 \Delta L_\beta
+ f_N \overline{N^c}i\tau_2\Delta N + \rm h.c \right] + \left[ Y\overline{N}\widetilde{H}\chi^0 
+ {\rm h.c.}\right]\,,
\end{equation}
where $L$ is the SM lepton doublet and $\alpha,\beta$ denote family indices. The Yukawa interactions importantly inherit the source 
of neutrino masses (terms in first square bracket) and DM-SM interactions (terms in second square bracket).

\subsection{Singlet-doublet fermion mixing}\label{singlet-doublet-mixing}
After electroweak phase transition vev of Higgs field introduces a mixing between $N^0$ and $\chi^0$.  The mass matrix is given by
\begin{equation}
\mathcal{M} = \begin{pmatrix} M_\chi & m_D\cr \\
m_D & M_N \,
\end{pmatrix}
\end{equation} 
where $m_D=Y v$. Diagonalizing the above mass matrix we get two mass eigenvalues:
\begin{eqnarray}
M_1 \approx M_\chi-\frac{m_D^2}{M_N-M_\chi}\nonumber\\
M_2 \approx M_N + \frac{m_D^2}{M_N-M_\chi}\label{mass-eigenstates}
\end{eqnarray}
where we have assumed $m_D << M_N, M_\chi$. The corresponding mass eigenstates are given by:
\begin{eqnarray}
N_1= \cos \theta \chi^0 + \sin \theta N^0\nonumber\\
N_2=\cos \theta N^0 - \sin \theta \chi^0\,,
\end{eqnarray}
where the mixing angle (in small mixing limit) is given by:
\begin{equation}\label{eq:sd_mix}
\tan 2\theta = \frac{2 m_D}{M_N-M_\chi}\,. 
\end{equation}
Due to the imposed $Z_2$ symmetry, the lightest odd particle remains stable. We choose $N_1$ to be the lightest and hence becomes 
a viable candidate for DM. The next-to-lightest $Z_2$ odd particle is charged lepton $N^{\pm}$ whose mass in terms of 
$M_1$, $M_2$ and mixing angle $\theta$ is given by:
\begin{equation}
M^\pm = M_1 \sin ^2 \theta + M_2 \cos^2 \theta \simeq M_N \,.
\end{equation}

From Eq.~\ref{eq:sd_mix}, we see that $Y$ and $\sin \theta$ are not two independent parameters. They are 
related by: 
\begin{equation}
Y = \frac{\Delta M \sin 2\theta} {2 v} \, ,
\label{eq:Y-DM}
\end{equation}
with $\Delta M = M_2 -M_1$. We use $\sin \theta$ as an independent parameter in our analysis. We will see that the 
mixing angle plays a vital role in the DM phenomenology. In particular, the relic abundance of DM gives an upper
bound on the singlet-doublet mixing angle to be $\sin \theta \lesssim 0.4$. For larger mixing angle the relic abundance is less 
than the observed value due to large annihilation cross-sections in almost all parameter space. We also found that a lower 
bound on $\sin \theta$ coming from the decay of $N_2$ and $N^-$ after they freeze out from the thermal bath. 
In principle these particles can decay on, before or after the DM ($N_1$) freezes out depending on the mixing angle. 
In the worst case,  $N_2$ and $N^-$ have to decay before the onset of Big-Bang nucleosynthesis. In that case, the lower bound 
on the mixing angle is very much relaxed and the out-of-equilibrium decay of $N_2$ and $N^-$ will produce an additional abundance 
of DM. Therefore, in what follows, we demand that $N_2$ and $N^-$ decay on or before the freeze out of DM ($N_1$). As a result we 
get a stronger lower bound on $\sin \theta$, which of course depends on their masses.

If the mass splitting between $N^-$ and $N_1$ is larger than $W^\pm$-boson mass, then $N^-$ decay preferably through the two body 
process: $N^- \to N_1 + W^-$. However, if the mass splitting between $N^-$ and $N_1$ is less than $W^\pm$-boson mass then 
$N^-$ decay through the three body process: $N^- \to N_1 \ell^- \overline{\nu_\ell}$. For the latter case, we get a 
stronger lower bound on the mixing angle than the former. The three body decay width of $N^-$ is given by \cite{Bhattacharya:2015qpa}:
\begin{equation}\label{N-decay}
\Gamma = \frac{ G_F^2 sin^2\theta}{24 \pi^3} M_N^5  I
\end{equation}
where $G_F$ is the Fermi coupling constant and $I$ is given as:
\begin{equation}\label{decay-rate}
I=\frac{1}{4}\lambda^{1/2}(1,a^2,b^2) F_1(a,b) + 6 F_2 (a,b)\ln  \left(\frac{2a}{1+a^2-b^2-\lambda^{1/2}(1,a^2,b^2)} \right) \,. 
\end{equation}
In the above Equation $F_1 (a,b)$ and $F_2 (a,b)$ are two polynomials of $a=M_1/M_N$ and $b=m_\ell/M_N$, where $m_\ell$ is the 
charged lepton mass. Up to ${\cal O}(b^2)$, these two polynomials are given by
\begin{eqnarray}
F_1 (a,b) &=& \left( a^6-2a^5-7a^4(1+b^2)+10a^3(b^2-2)+a^2(12b^2-7)+(3b^2-1)\right)\nonumber\\
F_2 (a,b) &=&  \left(a^5+a^4+a^3(1-2b^2)\right)\,.
\end{eqnarray} 
In Eq. (\ref{decay-rate}), $\lambda^{1/2}=\sqrt{1+a^4+b^4-2a^2-2b^2-2a^2b^2}$ defines the phase space. In the limit $b=m_\ell/M_N 
\to 1-a=\Delta M/M_N$, $\lambda^{1/2}$ goes to zero and hence $I\to 0$. The life time of $N^-$ is then given by 
$\tau \equiv \Gamma^{-1}$. We take the freeze out temperature of DM to be $T_f= M_1/ 20$. Since the DM freezes out during radiation 
dominated era, the corresponding time of DM freeze-out is given by :
\begin{equation}
t_f= 0.301 g_\star ^{-1/2} \frac{m_{\rm pl}} {T_f^2} \, ,
\end{equation}
where $g_\star$ is the effective massless degrees of freedom at a temperature $T_f$ and $m_{\rm pl}$ is the Planck mass. Demanding 
that $N^-$ should decay before the DM freezes out (i.e. $\tau \lesssim t_f$) we get 
\begin{equation}\label{theta_constraint}
\sin \theta \gtrsim 1.1789 \times 10^{-5} \, \, \left(\frac{1.375\times
  10^{-5}} {I} \right)^{1/2} \left(  \frac{200 \rm GeV }{M_N}
\right)^{5/2} \left( \frac{g_\star}{106.75} \right)^{1/4} \left (
  \frac{M_1} {180 \rm GeV}\right)\,.
\end{equation}
Notice that the lower bound on the mixing angle depends on the mass of $N^-$ and $N_1$. For a typical value of
$M_N=200$ GeV, $M_1=180$ GeV, we get $\sin \theta \gtrsim 1.17 \times 10^{-5}$. Since $\tau$ is inversely
proportional to $M_N^5$, larger the mass, smaller will be the lower bound on the mixing angle. We will come 
back to this issue while calculating the relic abundance of DM in section~\ref{relic}.

\subsection{Doublet-triplet scalar mixing }
In the scalar sector, the model constitutes an usual Higgs doublet and an additional triplet. The quantum fluctuations around the 
vacuum is given as: 
\begin{equation}
H^0 = \frac{1}{\sqrt{2}}(v + h^0 + i \xi^0 ) , \Delta^0 = \frac{1}{\sqrt{2}}(u_\Delta + \delta^0 + i \eta^0 )
\end{equation}
The mass matrix is given as :
\begin{equation}
\mathcal{M}_{sc}^2 = \begin{pmatrix} M_H^2 & \mu v/2 \cr \\
\mu v/2 & M_\Delta^ 2 \,
\end{pmatrix}
\end{equation} 
where $M_\Delta^2 = \mu_\Delta ^2 + \lambda_{H\Delta} v^2/2$. The two neutral Higgs fields (CP - even) mass eigenstates are given by
\begin{equation}
H_1 =\cos\theta_0 h^0 + \sin\theta_0 \delta^0 , H_2= -\sin \theta_0 h^0 + \cos \theta_0 \delta^0
\end{equation}
 where $H_1$ is the standard model like Higgs and $H_2$ is the triplet like Higgs. The mixing angle is given by
 \begin{equation}
 \tan 2\theta_0 = \frac{ \mu v}{(M_\Delta^2 - M_H^2)}\label{mix}\,.
 \end{equation}
The corresponding mass eigenvalues are $M_{H_1}$ (SM Higgs like) and $M_{H_2}$ (triplet like) and are given as :
 \begin{eqnarray}
 M_{H_1}^2 \approx M_H^2-\frac{(\mu v/2)^2}{M_\Delta^2-M_H^2}\nonumber\\
M_{H_2}^2 \approx M_\Delta^2+\frac{(\mu v/2)^2}{M_\Delta^2-M_H^2}\,.
 \end{eqnarray}
 
\begin{figure}[thb]
$$
\includegraphics[height=6.6cm]{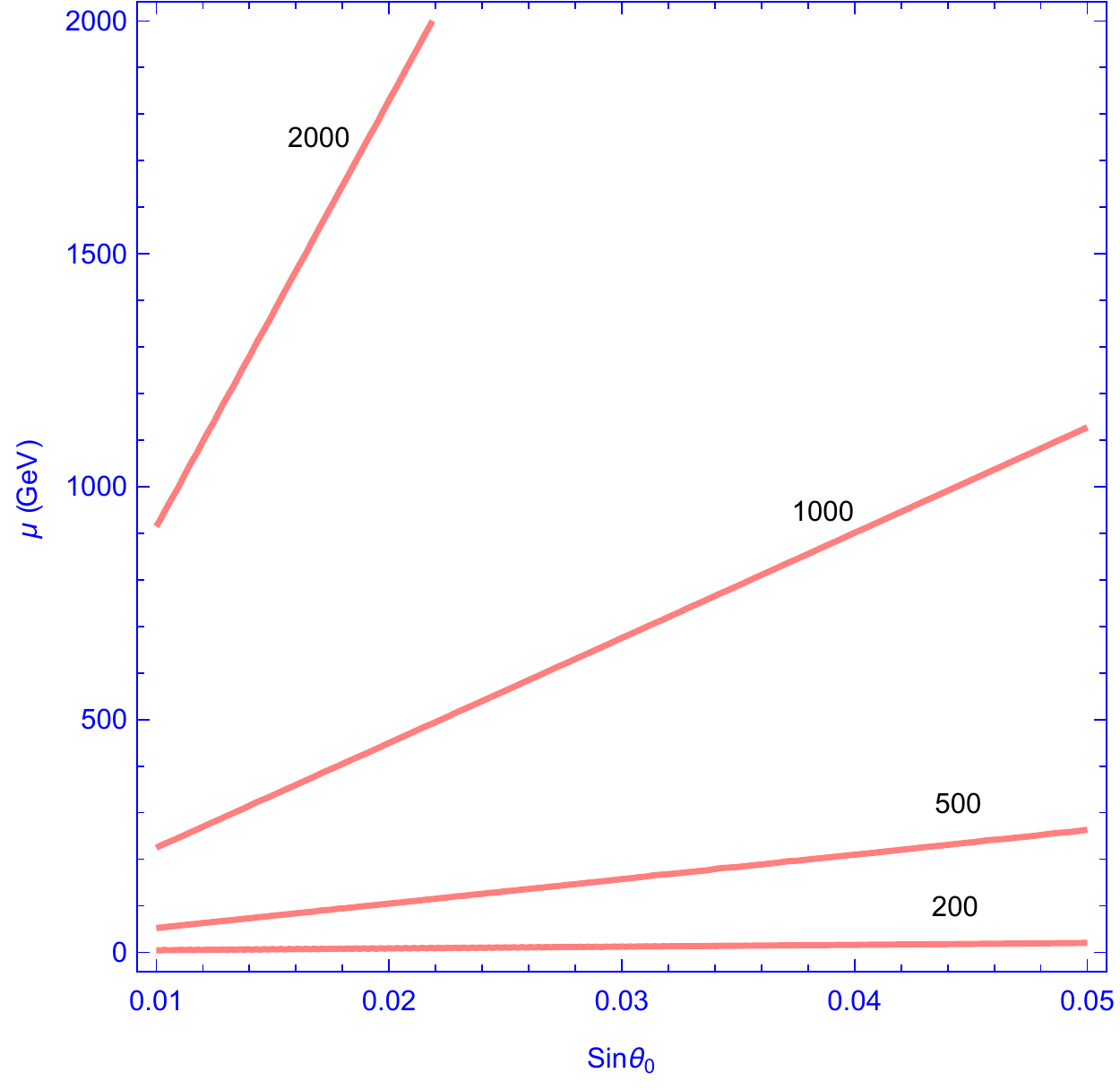}
$$
\caption{Contours of different values of $M_\Delta$ (in GeV) in the plane of $\mu$  versus $\sin \theta_0$.}
\label{contour}
\end{figure}

Since the addition of a scalar triplet can modify the $\rho$
parameter, which is not differing from SM value: $\rho=1.00037 \pm
0.00023$ \cite{pdg}, so we have a constraint on the vev $u_\Delta$ as:
\begin{equation}\label{const}
 u_\Delta  \leq 3.64 \rm GeV\,.
\end{equation}
For different values of $M_\Delta$ we have shown $\mu$ as a function of $\sin \theta_0$ in Fig. (\ref{contour}). 
Here we see that smaller is the triplet scalar mass, the smaller is the dependence on mixing angle $\sin\theta_0$.

From Eqs.(\ref{mix}), (\ref{vev}) and (\ref{const}) we see that there exist an upper bound on the mixing angle 
\begin{equation}
\sin \theta_0 < 0.02 \left( \frac{174 {\rm GeV} }{v} \right) \left(
  \frac{1}{1 - 0.39 \frac{(M_H/125 \rm GeV)^2}{(M_\Delta/200 \rm GeV)^2}}  \right)\,.
\end{equation}

We also get a constraint on $\sin \theta_0$ from SM Higgs phenomenology, since the mixing can change the strength of the Higgs
coupling to different SM particles. See for example~\cite {Cheung:2015dta, Hartling:2014aga}, in which the global 
fit yields a constraint on mixing angle $\sin \theta_0 \lesssim 0.5$, which is much larger than the constraint obtained 
using $\rho$ parameter. We may also get a constraint on $\sin \theta_0$ from the decay of SM Higgs to different channels. 
For example, let us take the decay of $H_1$ to $\tau$ leptons. The decay width is given by:
\begin{equation}
\Gamma = \frac{M_{H_1}}{8 \pi} \frac{m_\tau^2}{v^2} \left(1-\frac{4
    m_\tau^2}{M_{H_1}^2} \right)^{3/2} (1- \sin ^2 \theta_0)
\end{equation}
Comparing with the experimental branching fraction $Br(H_1 \rightarrow \tau \tau) = 6.272 \times 10^{-2}$, we 
found that $(\sin \theta _0)_{max} = 0.176$. So any value of the mixing angle less than this will be allowed by the 
corresponding decay width measurement. Similarly one can easily derive the limit on the doublet-triplet 
mixing from branching fraction of SM Higgs decaying to $W^+W^{-*},~ZZ^{*}$, which are much precisely measured at LHC. 
For example, if we choose Higgs decay to $W^+W^{-*}$ state, the observed branching fraction is $Br(H_1 \to W^+W^{-*}):~2.317\times
  10^{-1}$. In order to obtain a limit on the doublet-triplet mixing angle $\sin\theta_0$, we need to calculate the decay width of 
  this process process as given in~\cite{Keung:1984hn} :
\begin{equation}
\Gamma_{H_1\rightarrow WW^* \to Wf\bar{f^\prime}} = \frac{3 g^4 M_{H_1}}{512
  \pi^3} (g\sin\theta_0  u_\Delta /(4 M_w) - \cos\theta_0)^2 \, F(x) \,,
\end{equation}
where
\begin{equation*}
 F(x)=-|1-x^2| \left( \frac{47}{2} x^2 - \frac{13}{2} + \frac{1}
  {x^2} \right) + 3 (1-6x^2 +4 x^4) |Ln(x)| + \frac{3 (1-8x^2 +20
  x^4)} {|\sqrt{4x^2-1}|} \arccos \left[ \frac{3x^2-1} {2x^3} \right] \,,
\end{equation*}
with $x=M_W/M_{H_1}$. In the small mixing limit $\sin\theta_0 \rightarrow 0$, the decay reproduces same
branching ratio as that of the SM prediction. However, as we increase the value of the mixing angle, the branching ratio to this 
particular final state reduces due to larger triplet contributions. For example, with $\sin\theta_0=0.05,0.07,0.1$, $Br(H_1 \to WW^*)$  
is changed by $0.27\%, 0.51\%, 1.04\%$ respectively from the central value. Hence, in a conservative limit, if we take 
$\sin \theta_0 \sim 0.05$ or smaller, it is consistent with the experimental observation of Higgs decay to $WW^*$ final state.

Thus we see that the bound obtained on the mixing angle from Higgs decay depends on the final state that we choose, but is 
less constraining than that from the $\rho$ parameter. Therefore, we will use the constraint on the mixing angle, obtained from 
$\rho$ parameter, while calculating the DM-nucleon elastic scattering in section (\ref{dd}). Since the doublet-triplet scalar mixing 
is found to be small, we assume that the flavour eigenstates are the mass eigenstates and treat 
$M_{H_1}= M_H , M_{H_2}= M_\Delta$ throughout the calculation.

We also assume for simplicity that there is no mixing between the neutral CP-odd states and in the charged scalar states, so that 
$\xi^0$ is absorbed by the gauge bosons after spontaneous symmetry breaking in unitary gauge  and the charged triplet 
scalar fields remain as mass eigen states.

\section{Non Zero Neutrino Mass }\label{neutrino}   
The coupling of scalar triplet $\Delta$ to SM lepton and Higgs doublets combinely break the lepton number by two units as given in 
Eq. (\ref{yukawa_coupling}). As a result the $\Delta L_\alpha  L_\beta$ coupling yields Majorana masses to three flavors 
of active neutrinos as \cite{type_II}:
\begin{equation}\label{neutrino-mass}
(M_\nu)_{\alpha \beta}= \sqrt{2} (f_L)_{\alpha\beta}\langle \Delta \rangle \approx (f_L)_{\alpha\beta} \frac{-\mu v^2}{\sqrt{2} M_{\Delta}^2 }\,.
\end{equation}
Taking $\mu \simeq M_\Delta \simeq \mathcal{O}(10^{14})$ GeV, we can explain neutrino masses of order $0.1 \rm eV$ with a coupling strength 
$f_L \simeq 1$. However, the scale of $M_\Delta$ can be brought down to TeV scales by taking the smaller couplings.
 
To get the neutrino mass eigen values, the above mass matrix can be diagonalised  by the usual $U_{PMNS}$ matrix as :
\begin{equation}\label{eq:neu_mass}
M_\nu= U_{\text {PMNS}} \, M_\nu^{diag} \, U^T_{\text{ PMNS}} \,,
\end{equation}
where $U_{PMNS}$ is given by 
\begin{equation}\label{eq:pmns}
U_{\text PMNS} =
\begin{pmatrix}
c_{12}c_{13}  &  s_{12}c_{13}  & s_{13}e^{-i\delta_{13}}\\
-s_{12}c_{23}-c_{12}s_{23}s_{13}e^{i\delta_{13}} &
c_{12}c_{23}-s_{12}s_{23}s_{13}e^{i\delta_{13}}  & s_{23}c_{13} \\
s_{12}s_{23}-c_{12}c_{23}s_{13}e^{i\delta_{13}}  &
-c_{12}s_{23}-s_{12}c_{23}s_{13}e^{i\delta_{13}}  & c_{23}c_{13}
\end{pmatrix}
. U_{ph}\,,
\end{equation}
with $c_{ij}$, $s_{ij}$ stand for $\cos\theta_{ij}$ and
$\sin\theta_{ij}$ respectively and $U_{ph}$ is given by 
\begin{equation}
U_{ph}= \text {Diag} \left ( e^{-i\gamma_1} , e^{-i\gamma_2} , 1
\right ) \,.
\end{equation}
Where $\gamma_{1}$, $\gamma_2$ are two Majorana phases. The diagonal matrix $M_\nu^{diag}$ = Diag $(m_1,m_2,m_3)$
  with diagonal entries are
  the mass eigen values for the neutrinos.  The current neutrino oscillation data at $3\sigma$
  confidence level give the constraint on mixing angles ~\cite{pdg} :
\begin{equation}\label{eq:osc_angle}
0.259 < \sin^2\theta_{12} < 0.359,\,\,  0.374 <  \sin^2\theta_{23} <
0.628, \,\, 0.0176 <  \sin^2\theta_{13} < 0.0295
\end{equation}
However little information is available about the CP violating Dirac phase
$\delta$ as well as the Majorana phases.  Although the absolute mass
of neutrinos is not measured yet, the mass square differences have
already been measured to a good degree of accuracy :
\begin{eqnarray}\label{eq:mass_sqr}
\Delta m^2_0\equiv m_2^2-m_1^2= (6.99 - 8.18) \times 10^{-5} \rm eV^2 \nonumber \\
|\Delta m^2_{\rm atm} |\equiv |m_3^2-m_1^2|= (2.23 - 2.61) \times 10^{-3} \rm eV^2
\end{eqnarray}
One of the main issues of neutrino physics lies in the sign of the atmospheric mass
square difference $|\Delta m^2_{\rm atm} |\equiv |m_3^2-m_1^2|$, which is still unknown. This yields two possibilities: 
normal hierarchy (NH) ($m_1 < m_2 < m_3$) or inverted hierarchy (IH) ($m_3 < m_1 < m_2$). 
Another possibility, yet allowed, is to have a degenerate (DG) neutrino mass spectrum ($m_1 \sim m_2 \sim m_3$). 
Assuming that the neutrinos are Majorana, the mass matrix can be written as :
\begin{equation}\label{eq:mass_mat}
M_\nu = \begin{pmatrix}
 a &  b  & c \\
b &  d  & e \\
c &  e  & f
\end{pmatrix} 
\end{equation}
Using equations \ref{eq:neu_mass}, \ref{eq:pmns}, \ref{eq:osc_angle} and \ref{eq:mass_sqr},
we can estimate the unknown parameters in neutrino mass matrix of
Eq.~(\ref{eq:mass_mat}). To estimate the parameters in NH, we use the
best fit values of the oscillation parameters. For a typical value
of $m_1 =0.0001$ eV, we get the mass parameters (in eV) as :
\begin{eqnarray}
a=0.003833, \, b=0.00759, \, c=0.002691 \nonumber \\
d=0.023865,  \,e=0.02083, \, f=0.03038
\end{eqnarray}
Similarly for IH case, choosing $m_3 =0.001$ eV, we get the mass
parameters (in eV) as :
\begin{eqnarray}
a=0.0484, \, b=-0.00459, \, c=-0.00573 \nonumber \\
d=0.02893,  \,e=-0.02366, \, f=0.02303
\end{eqnarray}
In both the cases, we put the Dirac and Majorana phases to be zero for simplicity.

The mass of the scalar triplet can also be brought down to TeV
scale by choosing appropriate Yukawa coupling.  If the mass is order of a few
hundreds of GeV,  then it can give interesting dilepton signals in the
collider. See for example, \cite{Dilepton} for a detailed discussion regarding 
the dilepton signatures at collider. 

We would like to note that the presence of scalar triplet addresses the 
issue of generating neutrino masses as we discussed here, and has minor dependence on DM relic density. However, the scalar 
triplet plays a major role in the direct detection by forbidding Z-mediated DM-nucleon interaction and thereby increasing 
the limit on singlet-doublet mixing as we will discuss shortly.

\section{Pseudo-Dirac nature of dark matter}

\subsection{Pseudo-Dirac nature of Inert fermion doublet dark matter}
Let us assume the case where the singlet fermion $\chi^0$ is absent in the spectrum. In this case, the 
imposed $Z_2$ symmetry stabilizes the neutral component of the fermion doublet $N\equiv (N^0, N^-)^T$. 
From Eq. (\ref{yukawa_coupling}) we see that after EW phase transition the induced vev of the triplet yields a 
Majorana mass to $N^0$ and is given by: 
\begin{equation}\label{Majorana-mass}
m=\sqrt{2} f_N \langle \Delta \rangle \approx f_N \frac{-\mu v^2}{\sqrt{2} M_{\Delta}^2 }\,.
\end{equation} 
Thus the $N^0$ has a large Dirac mass $M_N$ as given in Eq. (\ref{Lagrangian}) and a small Majorana mass $m$ 
as shown in the above Eq. (\ref{Majorana-mass}). Therefore, we get a mass matrix in the basis $\{N^0_L, (N^0_R)^c\} $ as:
\begin{equation}
{\mathcal M} =
\begin{pmatrix}
  m  &  M_N \\
       M_N   & m
\end{pmatrix}
\end{equation}
The presence of small Majorana mass of the doublet DM splits the Dirac state $N^0$ into two pseudo-Dirac states: $\psi^0_{1,2}$, whose 
mass eigenvalues are given by $M_N \pm m$ for mixing angle $\pi /4$, which is the maximal mixing. Hence the mass splitting 
between the two states $\{N^0_L, (N^0_R)^c\} $ is:
\begin{equation}
\delta M= 2 m = 2 \sqrt{2} f_N u_\Delta\,.
\end{equation}
Notice that the above mass splitting $\delta M << M_N$ and hence does not play any role in the relic abundance calculation, where both 
the components act as degenerate DM components. However, the small mass splitting between the two pseudo-Dirac states prohibits $N_0$ to 
interact to the detector through $Z$ mediation in the non-relativistic inelastic scattering limit and is crucial to escape from the strong 
direct detection constraints mediated via $Z$-boson. For example, to explain the DAMA signal through the inelastic scattering of DM with 
the nuclei the required mass splitting should be $\mathcal{O} (100 \rm keV) $ ~\cite{Arina:2011cu,Arina:2012fb,Arina:2012aj}. 

A crucial observation from Eq. (\ref{neutrino-mass}) and (\ref{Majorana-mass}) is that the ratio:
\begin{equation}\label{coupling-ratio}
R=\frac{(M_\nu) } {m} = \frac{f_L} {f_N}
\end{equation} 
is extremely small. In particular, if $M_\nu \sim \mathcal{O} ( \rm eV)$ and $m \sim \mathcal{O} (100 {\rm KeV})$ then 
$R \sim 10^{-5} $. In other words the triplet scalar coupling to SM sector is highly suppressed in comparison to the 
DM sector. Using this constraint, in section \ref{relics_IF} we will calculate the relic abundance of inert fermion 
doublet DM.

\subsection{Pseudo-Dirac nature of singlet-doublet fermion dark matter}\label{PD-dark-matter}
Next we adhere to the actual scenario where DM is the lightest one among the mixed states of singlet and doublet fermions $\chi^0$ and $N^0$. 
As discussed in section (\ref{singlet-doublet-mixing}), the DM is assumed to be $N_1= \cos \theta \chi^0 + \sin \theta N^0$ 
with a Dirac mass $M_1$. However, from  Eq. (\ref{yukawa_coupling}) we see that the vev of $\Delta$ induces a Majorana mass to 
$N_1$ due to singlet-doublet mixing and is given by:
\begin{equation}\label{majorana-mass}
m_1=\sqrt{2} f_N \sin^2 \theta \langle \Delta \rangle \approx f_N \sin^2 \theta \frac{-\mu v^2}{\sqrt{2} M_{\Delta}^2 }\,.
\end{equation}    
Thus the Majorana mass $m_1$ splits the Dirac spinor $N_1$ into two pseudo-Dirac states $\psi_1^{a,b}$ with masses $M_1 \pm m_1$. The 
mass splitting between the two pseudo-Dirac states ($\psi_1^{a,b}$) is given by 
\begin{equation}
\delta M_1= 2 m_1 = 2\sqrt{2} f_N \sin^2 \theta u_\Delta 
\end{equation}
Note that again $\delta M_1 << M_1$ from the estimate of induced vev of the triplet and hence does not play any role in the 
relic abundance calculation. However, the sub-GeV order mass splitting plays a crucial role in direct detection by forbidding 
the Z-boson mediated DM-nucleon elastic scattering. We will come back to this issue while discussing the inelastic scattering 
of DM with nucleon in sec.~\ref{DDin}. Now from Eq. (\ref{neutrino-mass}) and (\ref{majorana-mass}) we see that the ratio: 
\begin{equation}\label{improved-coupling-ratio}
R=\frac{(M_\nu)} {m_1} = \frac{f_L} {f_N \sin^2 \theta}\,.
\end{equation}
Thus in comparison to Eq. (\ref{coupling-ratio}), we see that the ratio between the two couplings $R=f_L/f_N$ is improved 
by two orders of magnitude ({\it i.e.} $R \sim 10^{-3}$) if we assume $\sin \theta =0.1$, which is the rough order of magnitude 
of singlet-doublet mixing being used in relic abundance calculation as we demonstrate in the next section.

\section{Relic Abundance of DM}\label{relic}

\subsection{ Relics of Inert fermion doublet dark matter}\label{relics_IF}
In absence of the singlet fermion $\chi^0$, the neutral component ($N^0$) of the fermion doublet is stable 
due to the imposed $Z_2$ symmetry. However, this does not guarantee that the $N^0$ alone is a viable dark matter 
candidate. Under this circumstance it is crucial to check if $N^0$ can give rise correct relic abundance observed 
by WMAP and PLANCK. 

The relic abundance of a DM is characterised by the number changing processes in which the candidate is involved. In this case, 
on top of annihilations to SM particles, the DM ($N_0$) can also participate in co-annihilations with heavier particles $N_{\pm}$ which 
are odd under the same $Z_2$ symmetry. The relevant annihilation and co-annihilation channels in order to keep the inert fermion 
doublet DM in the thermal equilibrium in the early universe are listed below.\\

\noindent $N^0 \overline{N^0} \rightarrow HH,ZH, W^+W^-,ZZ, \Delta^{+ + } \Delta ^{--} ,
  \Delta^+ \Delta^-,
  \Delta^0 \Delta^0 , W^\mp \Delta^\pm ,\Delta^0 H, \Delta^0 Z
  ,f\bar{f}$ \\
 $N^0 N^\pm \rightarrow W^\pm \gamma, W^\pm H, W^\pm
 Z,\Delta^\pm Z, \Delta^\pm
 H, \Delta^\pm \gamma, W^\pm \Delta^0, \Delta^{\pm \pm} W^\mp,
 \Delta^0 \Delta^\pm, f^\prime \bar{f}$ \\
 $N^\pm N^\mp \rightarrow W^\pm W^\mp, ZH, \gamma Z, \gamma \gamma, Z Z, \Delta^{++} \Delta^{--} , \Delta^{+} \Delta^-,
W^+ \Delta^-, Z \Delta^0 ,f\bar{f}
$\\

\begin{figure}
  \includegraphics[width =0.6 \textwidth]{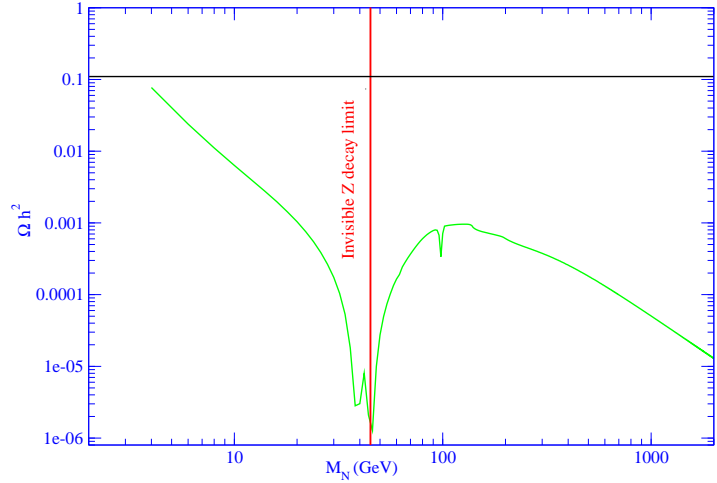}
\caption{Relic abundance (green line) of $N_0$, the neutral component of the doublet as DM, plotted as a function of 
doublet mass ($M_N$) in GeV. Black horizontal line shows the observed relic abundance by PLANCK data. The solid red vertical line is 
shown to mark $\frac{M_Z}{2}=45$ GeV; for $M_N>\frac{M_Z}{2}$ the DM does not contribute to the invisible Z decay width.}
\label{fig:dblt_dm}
\end{figure}

We use micrOMEGAs~\cite{micro} to calculate the relic abundance of dark matter. In Fig.~\ref{fig:dblt_dm}, we have shown 
the relic abundance of $N^0$ dark matter as a function of its mass. In a conservative limit we take the mass splitting between 
$N^0$ and its charged partner $N^-$ to be 1 GeV. The scalar triplet mass is fixed at 200 GeV and its coupling with the fermions 
is taken to be $\frac{f_L}{f_N} =10^{-5}$. We see that the large annihilation and co-annihilation cross-sections 
always yield much smaller relic density than required and hence the model is ruled out with the mass range of the order of TeV. The 
dominant channels are $N^0 \overline{N^0} \rightarrow HH,ZH, W^+W^-,ZZ$ and $N^\pm N^\mp \rightarrow W^\pm W^\mp$. We can also 
clearly spot the resonance at $M_N=\frac{M_Z}{2}$, where the relic density drops due to enhancement in the cross-section due to 
s-channel $Z$ mediation. The resonance drop at $M_N=100$ GeV specifies the presence of triplet. Thus we infer that the neutral 
component of the doublet alone can not be a viable DM candidate as its relic abundance is much below the observed limit. 
Therefore, in the next section we will consider a mixed singlet-doublet state as the candidate of DM.

\subsection{Relics of Singlet-Doublet mixed fermion dark matter}
The singlet ($\chi^0$) and neutral component of the doublet ($N^0$) fermion mix with each other after the electroweak 
symmetry breaking. In this scenario, the lightest particle $N_1= \cos \theta \chi^0+ \sin \theta N^0$, which is stabilized 
by the imposed $Z_2$ symmetry, serves as a viable candidate of DM. The relic abundance of $N_1$ can be obtained through 
its annihilations to SM particles as well as through co-annihilations with $N^-$ and $N_2$. 
The main processes which contribute to the relic abundance of DM without involving triplet scalar 
are \cite{Bhattacharya:2015qpa} :\\

\noindent $N_1 \overline{N_1} \rightarrow HH,ZH, W^+W^-,ZZ,f\bar{f} \\
 N_1 \overline{N_2} \rightarrow HH,ZH, W^+W^-,ZZ,f\bar{f} \\
N_2 \overline{N_2} \rightarrow HH,ZH, W^+W^-,ZZ,f\bar{f} \\
 N_1 N^\pm \rightarrow W^\pm \gamma, W^\pm H, W^\pm
 Z,f^\prime \bar{f} \\
N_2 N^\pm \rightarrow W^\pm \gamma, W^\pm H, W^\pm
 Z,f^\prime \bar{f} \\
 N^\pm N^\mp \rightarrow W^\pm W^\pm, ZH, \gamma Z, \gamma \gamma, Z Z, f\bar{f}
$\\

In presence of the light scalar triplet $\Delta$, there will be additional s-channel processes through $\Delta^0$ mediation 
as well as processes involving $\Delta$ particles in the final states. The relevant processes are :\\

\noindent $ N_1 \overline{N_1} \xrightarrow {\Delta^0} f\bar{f}, HH, W^+W^-,ZZ $\\
  $N_1 \overline{N_1} \rightarrow \Delta^{+ + } \Delta ^{--} ,
  \Delta^+ \Delta^-
  \Delta^0 \Delta^0 , W^\pm \Delta^\pm ,\Delta^0 H, \Delta^0 Z $\\
$N_1 \overline{N_2} \xrightarrow {\Delta^0} f\bar{f}, HH, W^+W^-,ZZ$ \\
 $ N_1 \overline{N_2} \rightarrow \Delta^{+ + } \Delta ^{--} ,
  \Delta^0 \Delta^0,
  \Delta^+ \Delta^-, W^\pm \Delta^\pm ,\Delta^0 H, \Delta^0 Z$ \\
$N_1 N^+ \rightarrow \Delta^- \Delta^{++}, W^- \Delta^{++}, \Delta^0
\Delta^+, H \Delta^+, Z \Delta^+, A \Delta^+, W^+ \Delta^0 $\\
$ N_2 \overline{N_2} \xrightarrow {\Delta^0} f\bar{f}, HH, W^+W^-,ZZ $\\
  $N_2 \overline{N_2} \rightarrow \Delta^{+ + } \Delta ^{--} ,
  \Delta^+ \Delta^-
  \Delta^0 \Delta^0 , W^\pm \Delta^\pm ,\Delta^0 H, \Delta^0 Z $\\
$N_2 N^+ \rightarrow \Delta^- \Delta^{++}, W^- \Delta^{++}, \Delta^0
\Delta^+, H \Delta^+, Z \Delta^+, A \Delta^+, W^+ \Delta^0 $\\
$N^\pm N^\pm \rightarrow \Delta^{++} \Delta^{--} , \Delta^{+} \Delta^-,
W^+ \Delta^-, Z \Delta^0$
\\

Relic density for $N_1$ is given by \cite{griest}
\begin{equation}
\Omega_{N_1} h^2 = \frac{1.09 \times 10^9 \rm Gev^{-1}}{g_\star^{1/2} m_{\rm pl}} \frac{1}{J(x_f)} ,
\end{equation}
where $J(x_f)$ is given by 
\begin{equation}
 J(x_f) = \int_{x_f}^\infty \frac{\langle \sigma |v| \rangle _{\rm eff}}{x^2} dx,
\end{equation}
where $\langle \sigma |v| \rangle _{\rm eff}$ is thermal average of annihilation and coannihilation cross-sections of the 
DM particle. The expression for effective cross-section can be written as :
\begin{equation}
 \begin{split}
  \langle \sigma |v| \rangle _{\rm eff}  =  & \frac{g_1^2}{g^2_{\rm eff}} \sigma (N_1 N_1) + 2 \frac{g_1 g_2}{g^2_{\rm eff}} \sigma (N_1 N_2)
  (1+\omega)^{3/2} exp(-x \omega) \\
  & + 2 \frac{g_1 g_3}{g^2_{\rm eff}} \sigma (N_1 N^-) (1+\omega)^{3/2} exp(-x \omega) \\
  & + 2 \frac{g_2 g_3}{g^2_{\rm eff}} \sigma (N_2 N^-) (1+\omega)^{3} exp(-2 x \omega) +
  \frac{g_2^2}{g^2_{\rm eff}} \sigma (N_2 N_2) (1+\omega)^{3} exp(-2 x \omega) \\
   & +  \frac{g_3^2}{g^2_{\rm eff}} \sigma (N^- N^-) (1+\omega)^{3} exp(-2 x \omega).
 \end{split}
\end{equation}

In this equation $g_1,g_2,g_3$ represent spin degrees of freedom for particles $N_1,N_2,N^-$ respectively and their
values are 2 for all. $\omega$ stands for the mass splitting ratio, given by $\omega = \frac{M_i-M_1}{M_1}$, where $M_i$ is the 
mass of $N_2$ and $N^\pm$. The  effective degrees of freedom denoted by $g_{\rm eff}$, and is given by
\begin{equation}
 g_{\rm eff} = g_1 + g_2(1+\omega)^{3/2} exp(-x \omega) + g_3 (1+\omega)^{3/2} exp(-x \omega)
\end{equation}

\begin{figure}[thb!]
$$
\includegraphics[height=6.5cm]{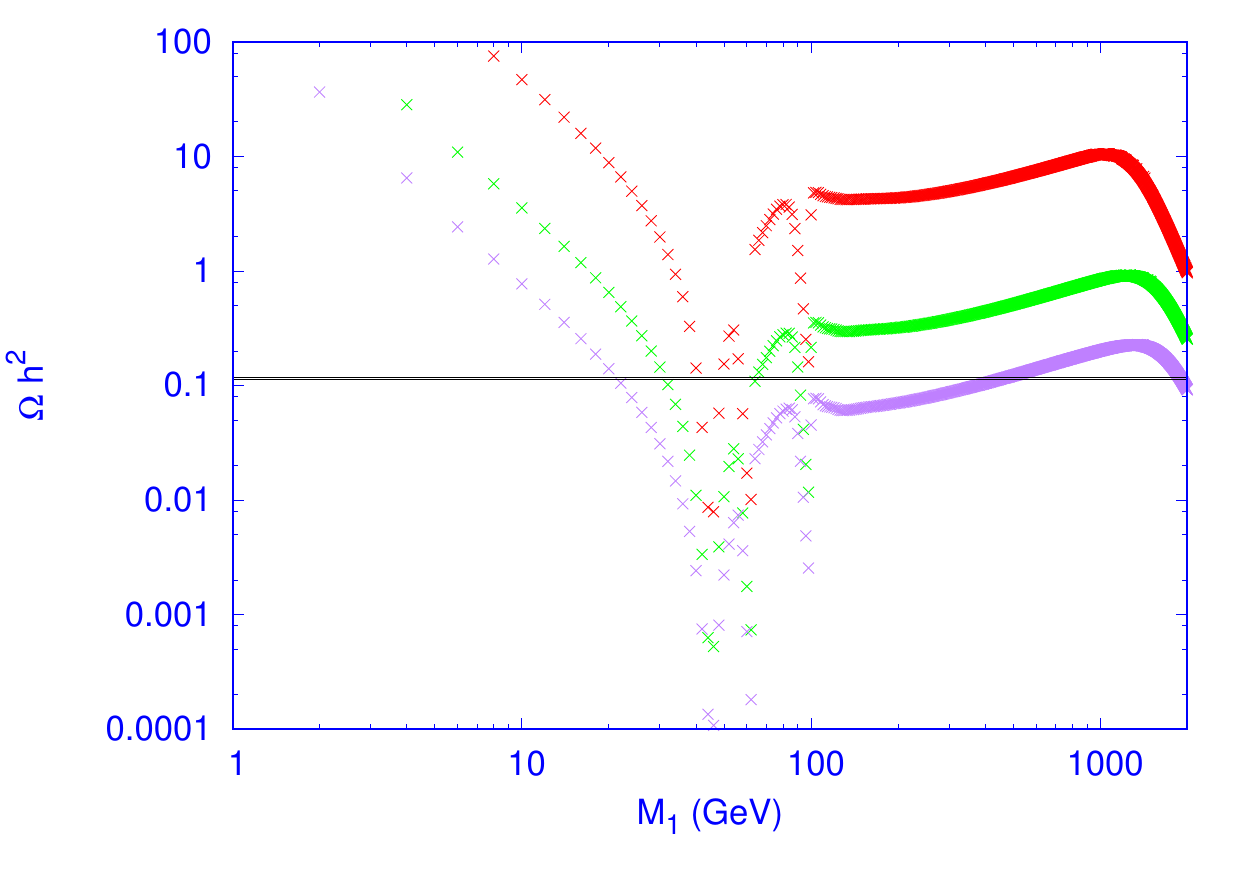}
\includegraphics[height=6.5cm]{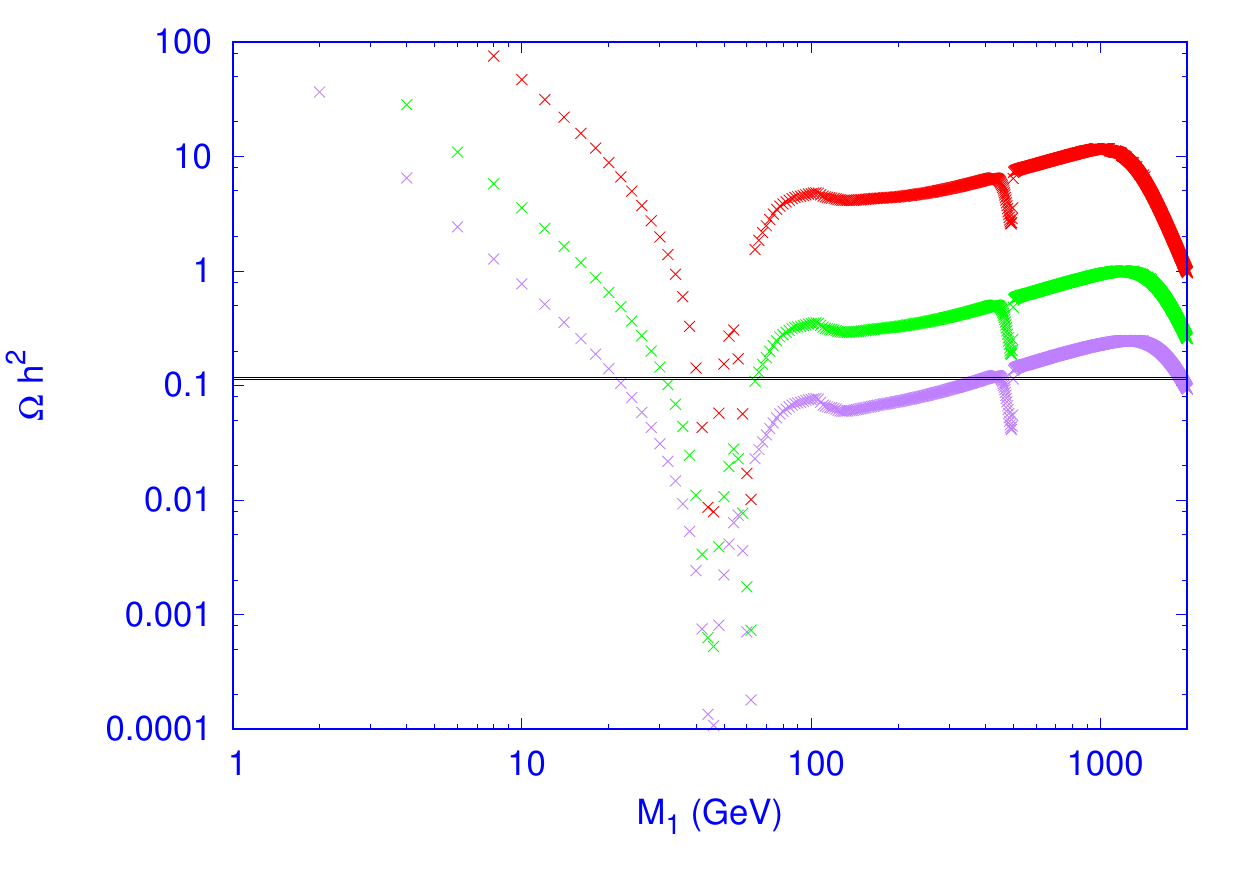}
$$
\caption{Relic density of DM as a function of its mass $M_1$ for different values of $\sin\theta = 0.1,0.2,0.3$, shown by 
        red (top), green (middle) and purple (bottom) respectively. The value of the triplet mass: $M_\Delta =200,1000 \rm GeV$ is fixed 
        respectively for left and right panel. All these plots are generated keeping a fixed value of the mass splitting $M_2-M_1 = 500$ GeV. 
        Ratio of Majorana couplings are fixed at : $\frac{f_L}{f_N}=10^{-3}$ for illustration.}
\label{omega_mass}
\end{figure}


To calculate the relic density of DM, we use the code micrOMEGAs~\cite{micro}. We have shown in fig. \ref{omega_mass} the relic
density as a function of DM mass keeping the mass difference fixed at $M_2-M_1=500$ GeV, for three different values of
the mixing angle: $\sin\theta= 0.1,0.2,0.3$, shown in red (top), green (middle), purple (bottom) respectively in the plot. In the left panel of the 
fig. \ref{omega_mass} we use $M_\Delta=200$ GeV, whereas in the right panel of fig. \ref{omega_mass} we use $M_\Delta=1000$ GeV. 
The black horizontal line corresponds to the observed relic density: $\Omega_{\rm DM}h^2 = 0.1199\pm 0.0027$ by PLANCK~\cite{PLANCK}. 
From fig. \ref{omega_mass}, we notice that there is a sharp decrease in relic density near three different points. These three 
points correspond to the resonant annihilation of DM to the SM particles via the s- channel processes mediated by $Z$ , h and $\Delta^0$. 
From these figures it is clear that as $\sin\theta$ increases relic density decreases. It is due to the fact that the $Z$ and $\Delta$ 
mediated cross-section increases for increase in $\sin\theta$, and hence yield a low relic density. For both the plots in 
fig~\ref{omega_mass} we fix the ratio of Majorana couplings to be: $\frac{f_L}{f_N}=10^{-3}$. From the plots in the fig. \ref{omega_mass}, 
we conclude that the $\Delta$ field is contributing to the relic density only near the resonance points. Apart from the resonance region, the 
triplet does not contribute significantly. This is because the total cross-section is dominated by $N_1\bar{N_1} \rightarrow W^+W-$ and the 
$\Delta$-mediated s-channel contribution is suppressed due to the large triplet scalar mass present in the propagator. Therefore, we can 
not expect any change in relic density allowed parameter space if we vary the ratio of Majorana couplings: 
$\frac{f_L}{f_N}$. The cross-sections involving scalar triplet in the final states also do not affect the 
relic abundance since those are suppressed by phase space due to heavy triplet masses and as in this region 
of parameter space ($M_1 > M_\Delta$) the cross-sections involving gauge bosons in the final state dominate. In summary, we don't 
see almost any difference in relic density of DM in left and right panel of Fig.~\ref{omega_mass} due to change in triplet masses. 
We can  however see that the resonance drop due to s-channel triplet mediation is reduced for large triplet mass 
$M_\Delta = 1000$ GeV (shown in right panel) in comparison to $M_\Delta =200$ GeV (shown in left panel) for obvious reasons. 
As the mass splitting between $N_1$ and $N_2$ is taken to be very large in the above cases, the dominant contribution to 
relic density comes from annihilation channels while co-annihilation channels are Boltzmann suppressed.

\begin{figure}[thb!]
$$
\includegraphics[height=6.0cm]{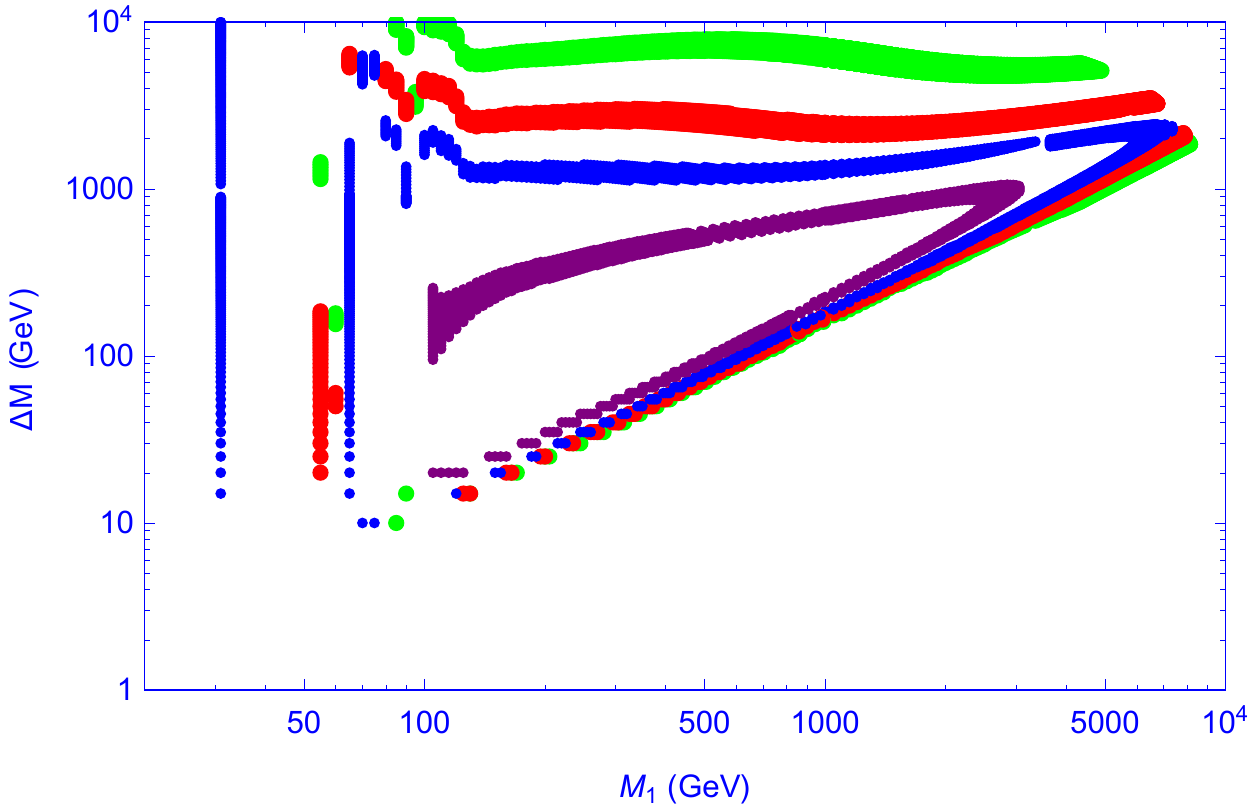}
\includegraphics[height=6.0cm]{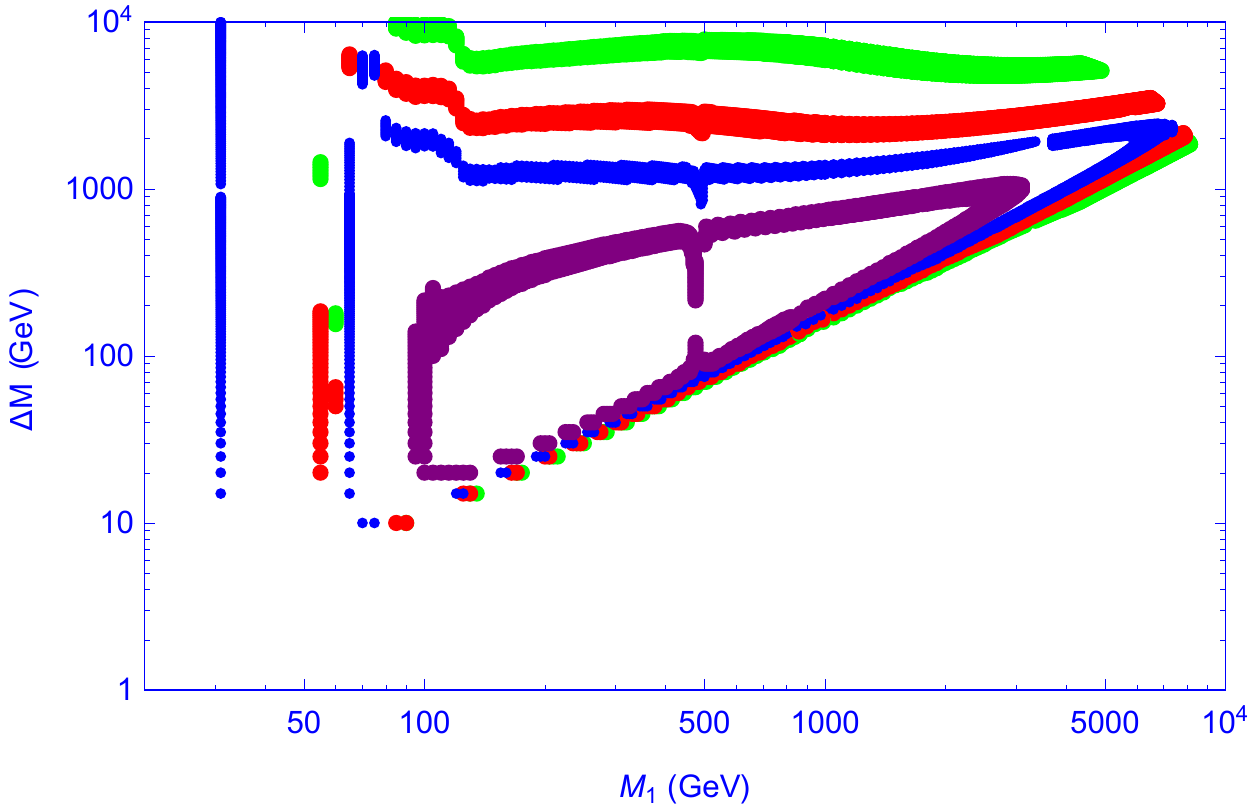}
$$
\caption{Scatter plot for correct relic density in the plane of $M_1$ and $\Delta M$, shown by green, 
         red, blue and purple coloured points for $\sin\theta=0.1,0.15,0.2,0.3$ respectively. Two different triplet masses 
         are chosen $M_\Delta = 200, ~\rm{and}~1000$ GeV respectively for the left and right panel plots. We fixed the value of Majorana coupling ratio: $f_L/ f_N =
         10^{-3}$ in both the figures for illustration.}
\label{omega_m1m2}
\end{figure}

Now we will show the effect of mass-splitting between $N_1$ and $N_2$ on DM relic density. In 
fig. \ref{omega_m1m2}, we have shown a scatter plot for correct relic density in the plane of $M_1$ and $\Delta M = M_2-M_1$. 
Green, red, blue and purple coloured points satisfy the constraint of relic density for $\sin\theta=0.1,0.15,0.2,0.3$ respectively 
(from outermost to innermost contour). Let us first consider the vertical bars in the left hand side of the allowed parameter space. 
In this region of small DM mass, annihilation cross-section achieves large enhancement due to s-channel $Z$ and $H$ mediation at 
$M_{N_1}=\frac{M_Z}{2}$ and at $M_{N_1}=\frac{M_H}{2}$ respectively where the annihilation cross-section is independent of $\Delta M$. 
Annihilation cross-sections contribute significantly for large $\Delta M$ to provide correct relic abundance. 
As the mass splitting decreases co-annihilation channels contribute significantly to add to the annihilation 
channels. As seen from the figure \ref{omega_m1m2}, we can divide the relic density allowed parameter space into two regions with 
same $\sin \theta $ value: i) The region in which $\Delta M$ is increasing with DM mass to satisfy correct 
relic density constraint. In this region, the contribution to relic density comes from both annihilation 
and dominantly from co-annihilation channels as the mass splitting is small. Here, due to small, $\Delta M$, the Yukawa coupling 
$Y$ (see Eq.~\ref{eq:Y-DM}) is small and so is the Higgs mediated cross-sections. Hence, co-annihilation channels 
provide with the rest of the requirement for correct relic density and allowed parameter space requires $\Delta M \sim M_1$.
ii) The second region corresponds to a large $\Delta M$ while insensitive to DM mass satisfying 
the correct relic abundance. In this region, the dominant contribution to relic density comes from the annihilation channels 
(large $\Delta M$ indicates large Yukawa $Y$ and large Higgs mediation cross-sections), and the co-annihilation channels are 
Boltzmann suppressed. $Z$ mediated annihilation cross-sections are fixed by the choice of a specific mixing angle (in the DM mass 
region within $\sim$ TeV). Therefore, the larger is the mixing the larger is the $Z$ mediated annihilation. This correctly balances 
the Higgs mediated annihilation cross-sections to yield correct relic density. That is why we notice that a smaller mass splitting ($\Delta M$) 
is required for larger $\sin\theta$ for a fixed value of DM mass. Hence green lines with smaller mixing ($\sin\theta=0.1$) requires 
larger $\Delta M$ and appears on top. With larger mixing, red, blue and purple lines, the required $\Delta M$ are smaller and appears below. 
It is easy to extend the analysis for even larger mixing angles, where the triangle becomes smaller and smaller in size and covers the 
innermost regions to yield the correct relic density. For $\sin \theta \gtrsim 0.5$ we can not get any relic abundance.  



Points below ``correct annihilation lines'' (for a specific value of $\sin \theta$) 
provide more than required annihilation and hence those are under abundant regions. Similarly just above those, the annihilation 
will not be enough to produce correct density and hence are over abundant regions. Points below (above) the ``correct co-annihilation regions'' 
produce more (less) co-annihilations than required and hence depict under (over) abundant regions. There is not much difference in 
the parameter space if we vary the scalar triplet mass except few points in the resonance region. It can be clearly seen in left 
and right panel of the fig.~\ref{omega_m1m2} with scalar triplet mass 200 GeV and 1000 GeV respectively. The Yukawa coupling ratio 
$f_L/f_N =10^{-3}$ is fixed for both the plots. Again, if we change this ratio to a different value, no significant change in the 
allowed parameter space is expected.


\begin{figure}[thb!]
$$
\includegraphics[height=6.5cm]{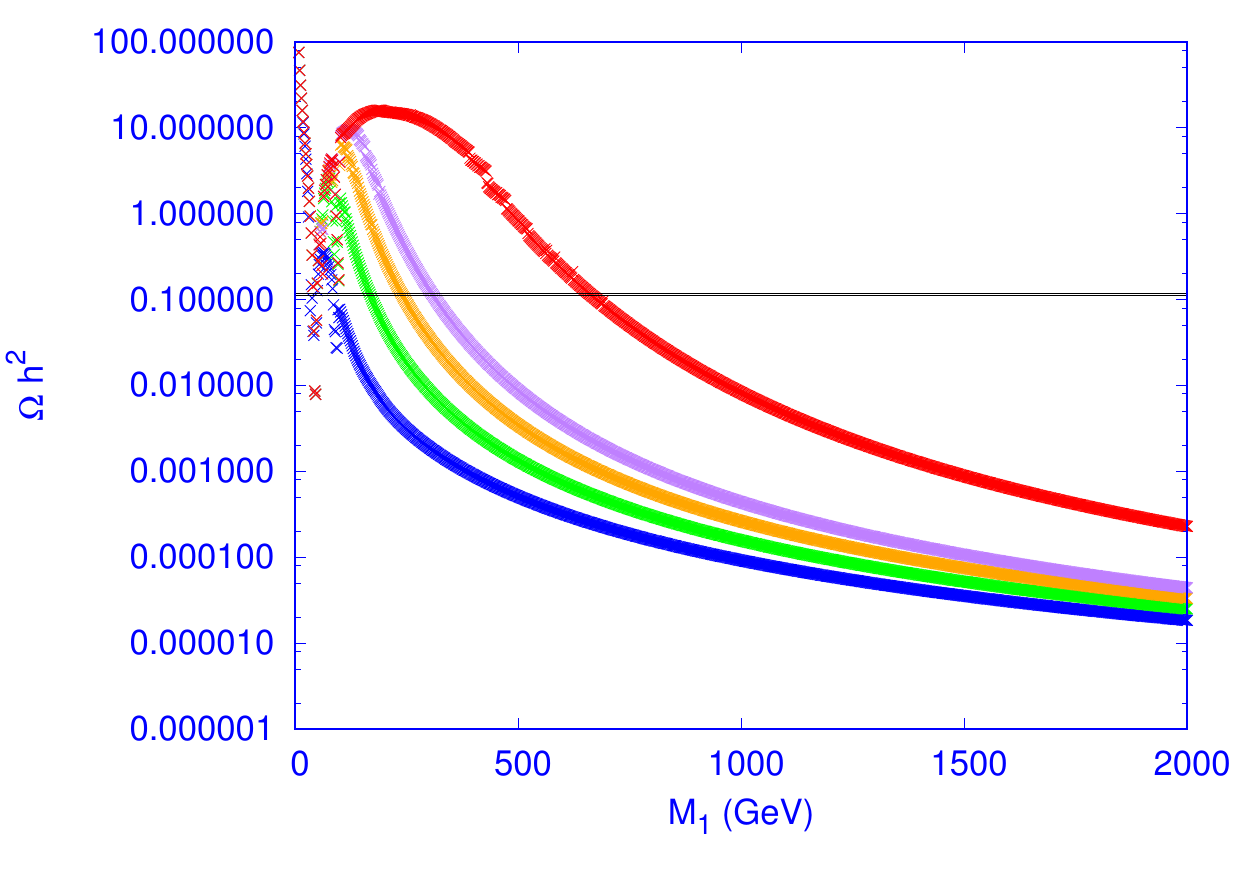}
\includegraphics[height=6.5cm]{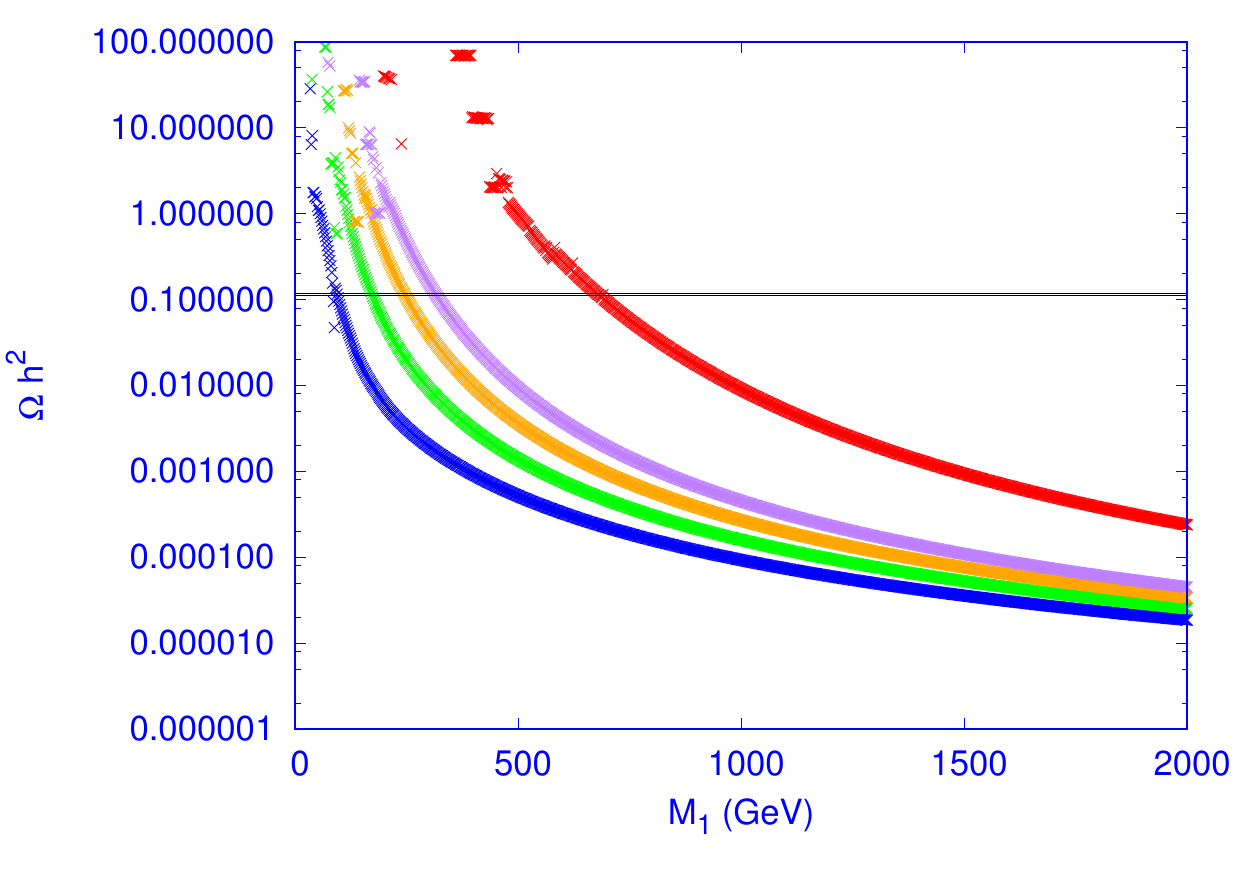}
$$
\caption{Left : $\Omega~h^2$ versus DM mass $M_{DM}$ in GeV for
          $\sin \theta =0.1$ and  $\Delta M= {10, 20,
   30, 40, 100}$ GeV (Blue, Green, Orange, Purple, Red respectively from bottom to top).
 Right :  $\Omega~h^2$ versus DM mass $M_{DM}$ in GeV for
          $\sin \theta =0.0001$ and  $\Delta M= {10, 20,
   30, 40, 100}$ GeV (Blue, Green, Orange, Purple, Red
 respectively from bottom to top). Horizontal line shows the correct relic density. We
 fixed the value $f_L/ f_N = 10^{-3}$ and $M_\Delta =200$ GeV
        for all the plots.}
\label{sin}
\end{figure}

The $\Delta M$ dependency on the relic density for a specific choice of mixing angle is shown in Fig. \ref{sin}, 
particularly for small mixing regions where co-annihilations play a crucial role in yielding correct relic density. In 
the left panel we use $\sin \theta=0.1$ and that in the right panel $\sin \theta=0.0001$. We plot different slices 
with constant $\Delta M = {10, 20, 30, 40, 100}$ GeV as shown in Blue, Green, Orange, Purple, Red respectively from bottom to top. We note 
here that with larger $\Delta M$, the annihilation cross-section increases due to enhancement in Yukawa coupling 
$Y \propto \Delta M$. However, co-annihilation decreases due to increase in $\Delta M$ as $\sigma \propto e^{- \Delta M}$. 
Note that in the small $\sin \theta$ limit the dominant contribution to relic density comes from the channels involving only 
$N_2$ and $N^\pm$ in the initial state going to SM gauge bosons, as mentioned in the beginning of this section. The processes 
involving $N_1 N_1 \to {\rm SM}$ are heavily suppressed with small $\sin \theta$. As a result, we first get relics of 
$N_2$ and $N^-$ which subsequently decay to $N_1$ before $N_1$ freezes out. In particular, if the mass splitting between $N^-$ and $N_1$ 
is more than 80 GeV, then $N^-$ decays through two body process: $N^- \to N_1 + W^-$. However, if the mass splitting between $N^-$ and 
$N_1$ is less than 80 GeV, than the former decays through the three body process, say $N^- \to N_1 +\ell^- + \overline{\nu_{\ell}}$.
Notice that the mixing angles $\sin \theta =0.1, 0.0001$ used simultaneously in the left and right-panel of Fig. (\ref{sin}) are 
much larger than the lower bound obtained on the singlet-doublet fermion mixing angle as given in eq.~\ref{theta_constraint} by considering 
the 3-body decay of $N^-$, namely $\sin \theta > \mathcal{O} (10^{-5})$. 

For large $\Delta M$ the co-annihilation cross-sections decrease, which are the dominant processes in the small 
$\sin \theta$ limit. As a result relic abundance increases for a particular value of $M_1$ with larger $\Delta M$. Hence we require a larger 
mass difference $\Delta M$ for larger DM mass to account for correct co-annihilation so that the relic density will be in the observed limit.

\section{ Direct Detection of DM through inelastic scattering with the nuclei}\label{DDin}
As discussed in section (\ref{relics_IF}), the inert fermion doublet $N^0$ alone does not produce correct relic 
abundance. Therefore, we refrain ourselves to consider the inelastic scattering of $N^0$ only with the nuclei mediated via 
$Z$ boson. Rather we will consider the inelastic scattering of DM $N_1$, which is an admixture of doublet $N^0$ and singlet 
$\chi^0$.  

From Eq. (\ref{Lagrangian}), the relevant interaction for scattering of $N_1$ with nucleon mediated 
via the Z-boson is given by 
\begin{equation}
{\mathcal L}_{\rm Z-DM} \supset \overline{N_1}\left( \gamma^\mu \partial_\mu + i g_z \gamma^\mu Z_\mu \right) N_1\,,
\end{equation}
where $g_z=\frac{g}{2 \cos \theta_w}\sin^2 \theta $. 
However the presence of scalar triplet, as discussed in section (\ref{PD-dark-matter}), splits the Dirac state $N_1$ 
into two pseudo-Dirac states  $\psi_1^{a,b}$ with a small mass splitting $m_1$. Therefore, the above interaction 
in terms of the new eigenstates $\psi_1^{a,b}$ can be rewritten as: 
\begin{equation}\label{eq: dm}
{\mathcal L}_{\rm Z-DM} \supset \overline{\psi_1^a} i\gamma^\mu \partial_\mu \psi_1^a + \overline{\psi_1^b} i\gamma^\mu \partial_\mu \psi_1^b 
+ i g_z \overline{\psi_1^a} \gamma^\mu  \psi_1^b Z_\mu \,.
\end{equation}

From the above expression the dominant gauge interaction is off-diagonal, and the diagonal interaction 
vanishes. As a result there will be inelastic scattering possible for the DM with the nucleus. Note that the mass splitting between 
the two mass eigen states $\psi_1^{a,b}$ is given by: $\delta M_1 = 2 \sqrt{2} f_N \sin^2\theta \, u_\Delta\,$. In this case, the minimum velocity of 
the DM needed to register a recoil inside the detector is given by \cite{TuckerSmith:2001hy,Cui:2009xq,Arina:2011cu,Arina:2012fb,Arina:2012aj} :
\begin{equation}
v_{\rm min}= c\sqrt{\frac{1}{2 m_n E_R}} \left( \frac{m_n E_R}{\mu_r}
  +  \delta M_1 \right)\,,
\end{equation}    
where $E_R$ is the recoil energy of the nucleon and $\mu_r$ is the reduced mass. If the mass splitting is above a few hundred keV, 
then it will be difficult to excite $\psi_1^b$ with the largest possible kinetic energy of the DM $\psi_1^a$. So the inelastic scattering 
mediated by $Z$-boson will be forbidden. As a result constraints coming from direct detection can be relaxed significantly. This in an important 
consequence in presence of the scalar triplet $\Delta$ in this model, which makes a sharp distinction with the existing analysis in this 
direction~\cite{Bhattacharya:2015qpa}.

\section{Direct Detection of DM through elastic scattering with the nuclei}\label{dd}
\begin{figure}[thb]
$$
\includegraphics[height=5.5cm]{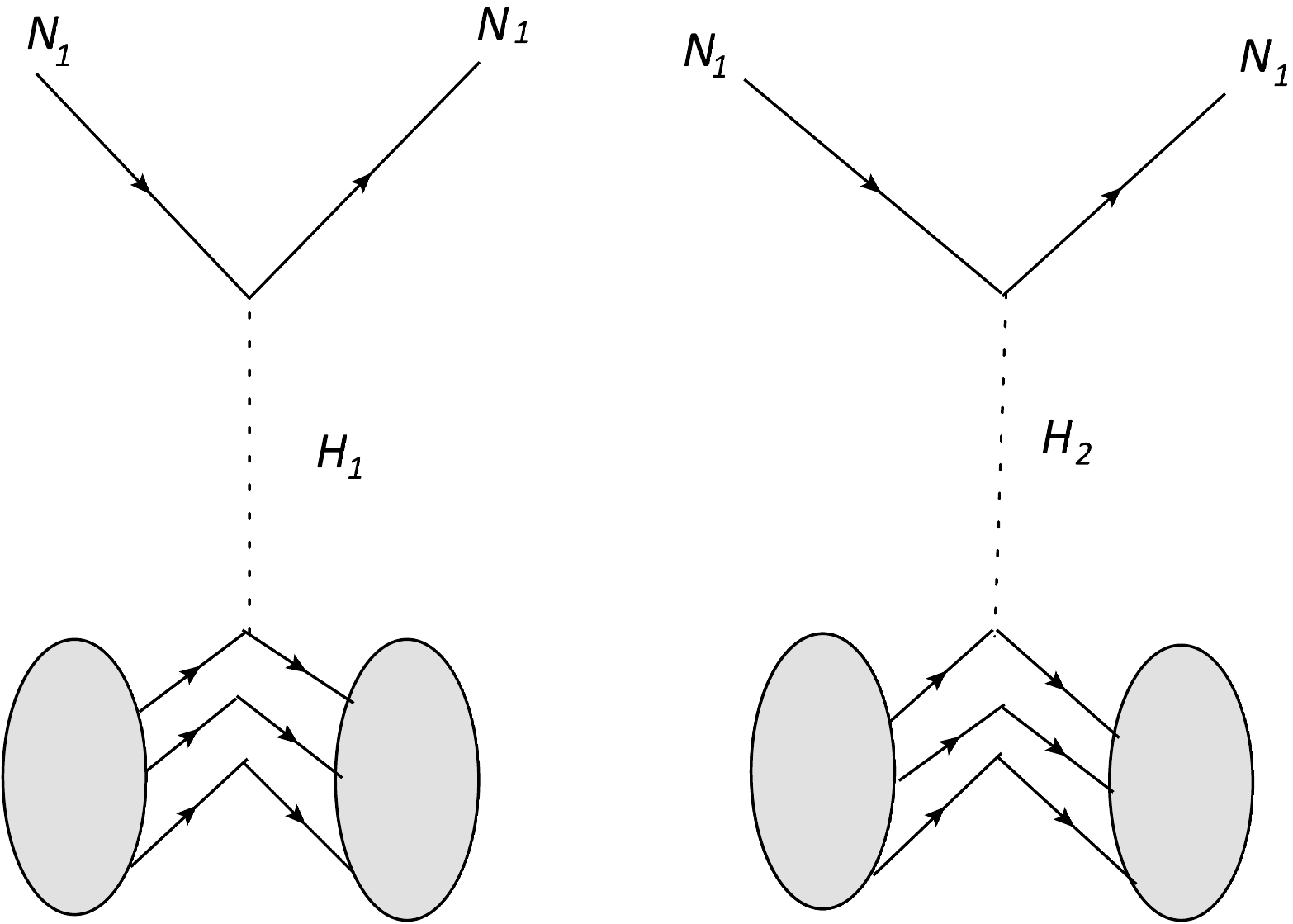}
$$
\caption{Feynman diagrams for direct detection of $N_1$ DM via Higgs mediation.}
\label{fig:DD2}
\end{figure}
We shall now point out constraints on the model parameters from direct search of DM via Higgs mediation. The relevant diagrams through 
which $N_1$ interacts with the nuclei are shown in Fig. (\ref{fig:DD2}). In particular, our focus will be on Xenon-100~\cite{xenon100} and 
LUX~\cite{Akerib:2016vxi} which at present give strongest constraint on spin-independent DM-nucleon cross-section from the null 
detection of DM yet. In our model, this in turn puts a stringent constraint on the singlet-doublet mixing angle $\sin \theta$ for 
spin independent DM-nucleon interaction mediated via the $H_1$ and $H_2$-bosons (see in the  Fig. (\ref{fig:DD2})). The cross-section per 
nucleon is given by~\cite{ Goodman:1984dc,Essig:2007az}
\begin{equation}\label{DM-nucleon-Z}
\sigma_{\rm SI} = \frac{1}{\pi A^2 }\mu_r^2 |\mathcal{M}|^2   
\end{equation}
where $A$ is the mass number of the target nucleus, $\mu_r=M_1 m_n/(M_1 + m_n) \approx m_n$ is the reduced mass, $m_n$ is the mass of nucleon 
(proton or neutron) and $\mathcal{M}$ is the amplitude for DM-nucleon cross-section.  There are two t-channel processes through which DM can 
interact with the nucleus which is shown in the fig \ref{fig:DD2}. The amplitude is given by: 
\begin{eqnarray}\label{scalar_mediated_crossssection}
\mathcal{M}=\sum_{i=1,2} \left[ Z f_p^i + (A-Z)f_n^i \right] 
\end{eqnarray} 
where the effective interaction strengths of DM with proton and neutron are given by:
\begin{equation}\label{f-values}
f_{p,n}^i = \sum_{q=u,d,s}f_{Tq}^{(p.n)} \alpha_q^i \frac{m_{(p,n)}}{m_q} + \frac{2}{27} f_{TG}^{(p,n)}\sum_{q=c,t,b} \alpha_q^i \frac{m_{p.n}}{m_q}
\end{equation}
with 
\begin{eqnarray}\label{alpha-value}
\alpha_q^1 = \frac{ Y\sin 2\theta \cos^2 \theta_0}{M_H^2} \left( \frac{m_q}{v}\right)
  \,\\
\alpha_q^2 = -\frac{ Y\sin 2\theta \sin^2 \theta_0}{M_\Delta^2} \left(
  \frac{m_q}{v}\right) \,.  \,
\end{eqnarray}
In Eq. (\ref{f-values}), the different coupling strengths between DM and light quarks are given by~\cite{DM_review1} $f^{(p)}_{Tu}=0.020\pm 0.004$, 
$f^{(p)}_{Td}=0.026\pm 0.005$,$f^{(p)}_{Ts}=0.118\pm 0.062$, $f^{(n)}_{Tu}=0.014\pm 0.004$,$f^{(n)}_{Td}=0.036\pm 0.008$,$f^{(n)}_{Ts}=0.118\pm 0.062$. 
The coupling of DM with the gluons in target nuclei is parameterized by 
\begin{equation}\label{Gluon-interaction}
f^{(p,n)}_{TG}=1-\sum_{q=u,,d,s}f^{(p,n)}_{Tq}\,. 
\end{equation}

\begin{figure}[htb!] 
$$
\includegraphics[height=5.8cm]{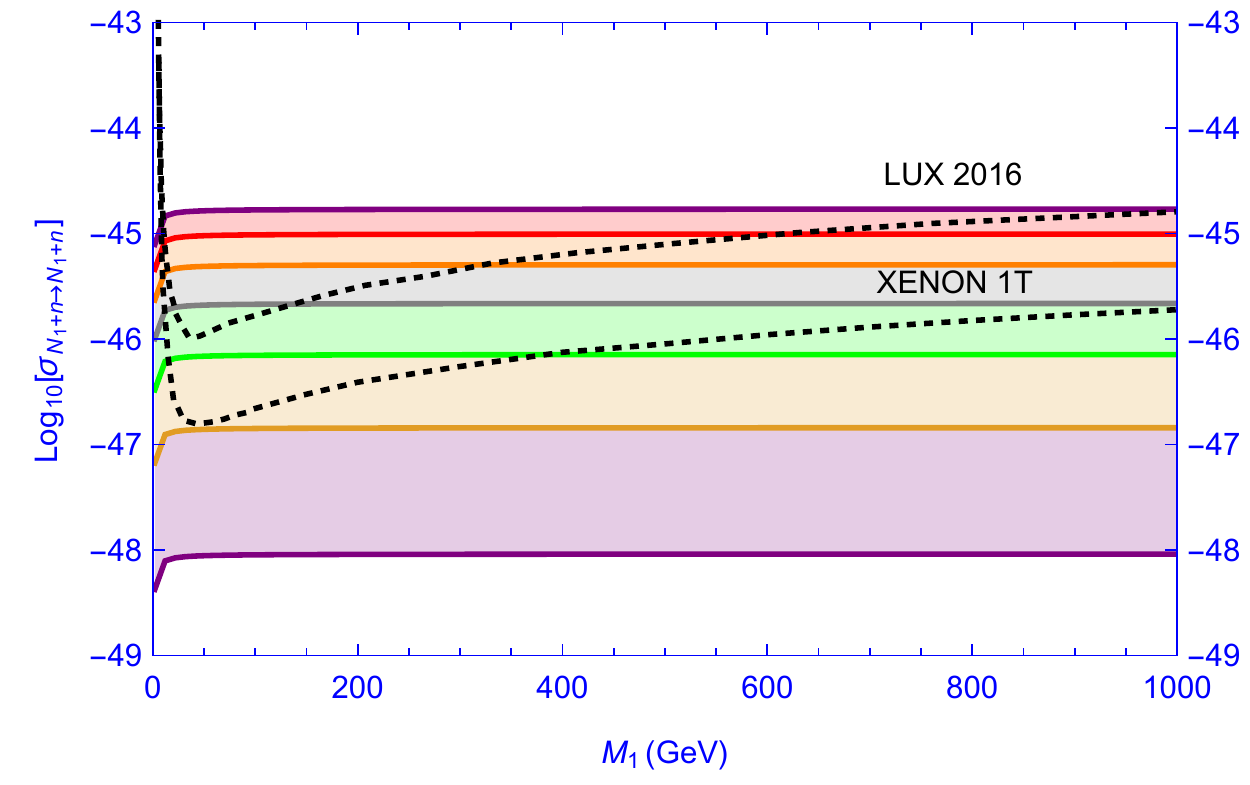}
\includegraphics[height=5.8cm]{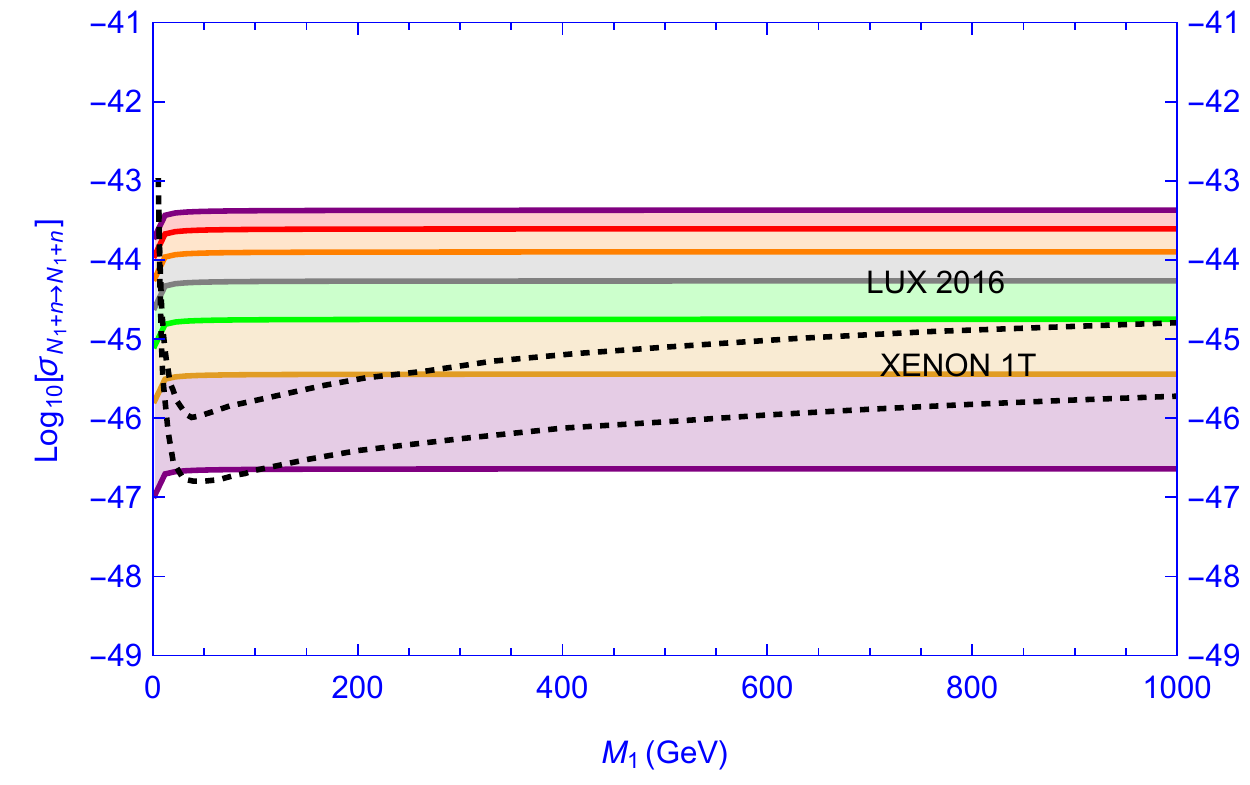}
$$
	\caption{Spin Independent direct detection cross-section for
          DM as a function of DM mass for $\sin\theta=$\{0.05-0.1\} (Purple), $\sin\theta=$\{0.1-0.15\} (Pitch), $\sin\theta =$\{0.15-0.2\} (Green), $\sin\theta
=$\{0.2-0.25\} (Gray), $\sin\theta =$\{0.25-0.3\} (Orange), $\sin\theta
=$\{0.3-0.35\} (Red). Black dotted curves show the data from  LUX and XENON 1T
prediction. Value of $\Delta M =100, ~500 ~\rm GeV$ are fixed for left and
right panel figures respectively. The scalar triplet mass is fixed at $M_\Delta =200 ~\rm GeV$ and scalar mixing angle is fixed at $\sin \theta_0 =0.05$  for the calculation.} 
\label{fig:DDcon}
\end{figure}

We have plotted the spin independent direct detection cross-section as a function of DM mass in the Fig.\ref{fig:DDcon} by 
taking the value of $M_\Delta =200 ~\rm GeV$ for two different values
of $M_2-M_1=100,~500 \rm ~GeV$ in the left and right panel respectively. 
The plot is generated using different values of the singlet-doublet mixing angle: 
$\sin\theta=$\{0.05-0.1\} (Purple), $\sin\theta=$\{0.1-0.15\} (Pitch), $\sin\theta =$\{0.15-0.2\} (Green), $\sin\theta
=$\{0.2-0.25\} (Gray), $\sin\theta =$\{0.25-0.3\}(Orange), $\sin\theta=$\{0.3-0.35\}(Red). 
The top Black dotted line shows the experimental limit on the SI nuclei-DM
cross-section with DM mass predicted from LUX 2016 and the one below shows the sensitivity of XENON1T. 
The constraint from XENON 100 is loose and weaker than the LUX data and hence not shown in the figure. One of the main outcome
 of the figure in the left panel is that with larger $\sin\theta$, due to larger Yukawa coupling direct search cross-section through 
Higgs mediation is larger. Hence, LUX data constrains the singlet-doublet mixing to $\sin\theta\sim 0.3$ for DM mass $\sim$ 600 GeV with $\Delta M= 100 ~\rm GeV$ (on the left hand side of Fig.~\ref{fig:DDcon}). The constraint on the mixing is even more weaker for larger DM mass $\sim 900$ GeV and can be as large as $\sin\theta\sim 0.4$. This 
presents a strikingly different outcome than what we obtained in absence of scalar triplet \cite{Bhattacharya:2015qpa}, the mixing angle 
was constrained there significantly to $\sin\theta\le 0.1$ to account for the null observation in LUX data. Again this is simply due to 
the absence of $Z$ mediated direct search processes due to the mass splitting generated by the triplet as discussed in the above section 
and hence allows the DM to live in a much larger region of relic density allowed parameter space. 
In the right panel of the Fig. \ref{fig:DDcon} with larger $\Delta M= 500 ~\rm GeV$, the constraint on $\sin \theta $ 
is more stringent than the left one. It is because
  the SI cross-section is enhanced due to the increase in Yukawa coupling
$Y \propto \Delta M$ for larger $\Delta M$ as expected. In the right panel, for DM mass of $\sim$ 300 GeV: 
$\sin\theta\sim 0.1$ and for DM mass around $\sim$ 1000 GeV and above: $\sin\theta\sim 0.15$ can be accommodated. 
Since the mixing between $\Delta - h$ is small: $\sin\theta_0 < 5\times 10^{-2}$ , the contribution to the cross-section by 
the $H_2$ mediated diagram is suppressed. This is also further suppressed by the large mass of $M_\Delta$ present in the 
propagator. For this reason no striking difference in direct search cross-section for higher values of $M_\Delta$ is found 
as the cross-section is dominated by $H_1$ mediation only.

\section{Decay of $N^-$ and the displaced vertex signature }\label{dec}
\begin{figure}[thb]
$$
\includegraphics[height=4.5cm]{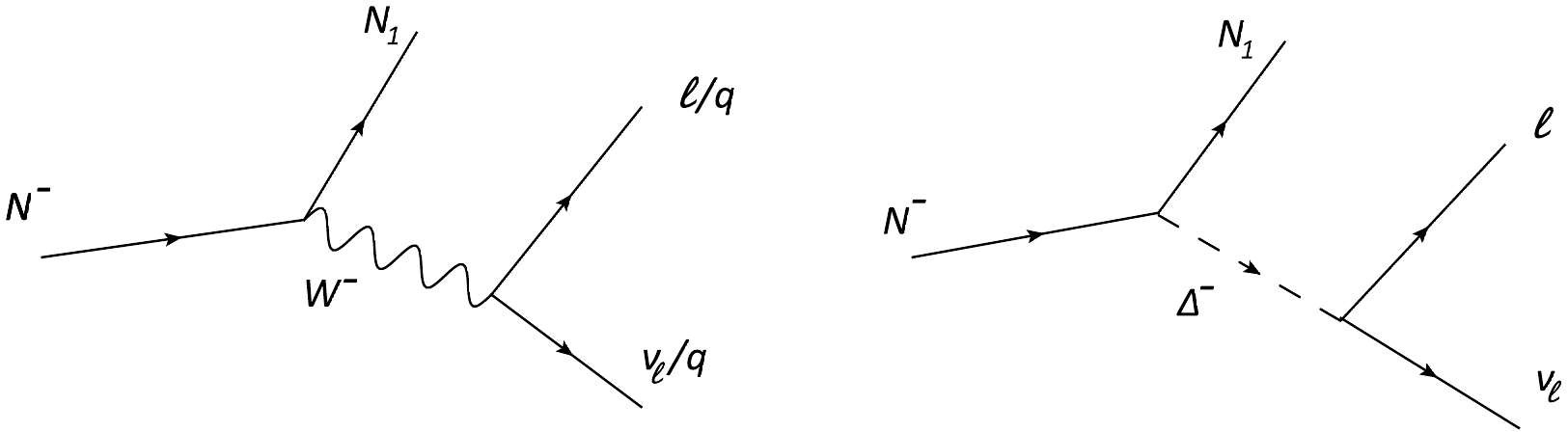}
$$
\caption{Feynman diagrams for three body decay of $N^-$ to DM.}
\label{fig:Decay}
\end{figure}

The phenomenology of the charged partner of DM is quite interesting. If the mass splitting between $N^\pm$ and $N_1$ is 
less than  mass of $W^-$ , then $N^-$ will decay via three body suppressed process: $N^- \to N_1 \ell \nu_\ell$ and 
$N^- \to N_1 + {\rm di-jets}$.  The relevant Feynman diagrams for the decay is shown in Fig. \ref{fig:Decay}. However the 
figure on the right side, mediated by triplet ($\Delta^-$), is suppressed due to the small coupling of triplet  
with the leptons and the large mass $M_\Delta$ present in the propagator.  
So the dominant contribution for decay of $N^-$ is coming from the left diagram of Fig. \ref{fig:Decay} through $W$ mediation. 
The decay rate for the process $N^- \to N_1 \ell \nu_\ell$ is given in Eq.\ref{N-decay}.

\begin{figure}[htb]
$$
      \includegraphics[width=.50\textwidth]{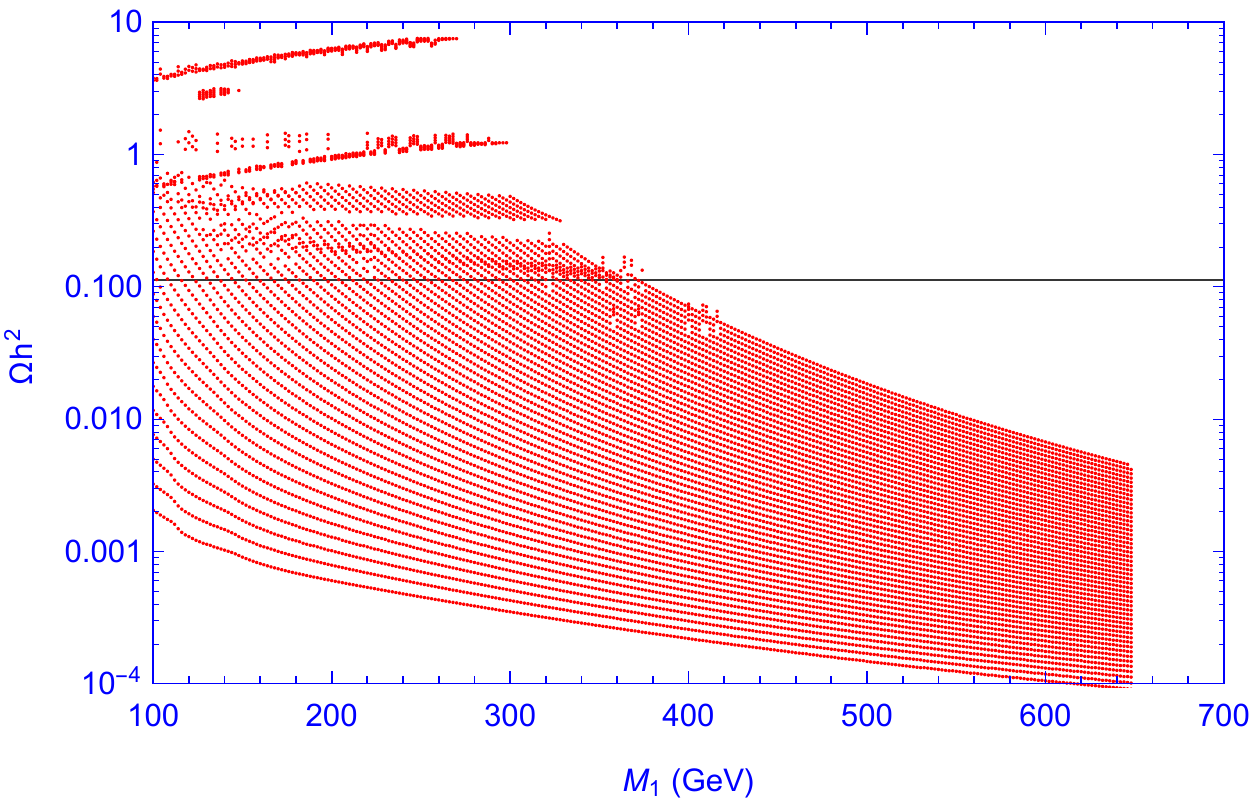}
      \includegraphics[width=.50\textwidth]{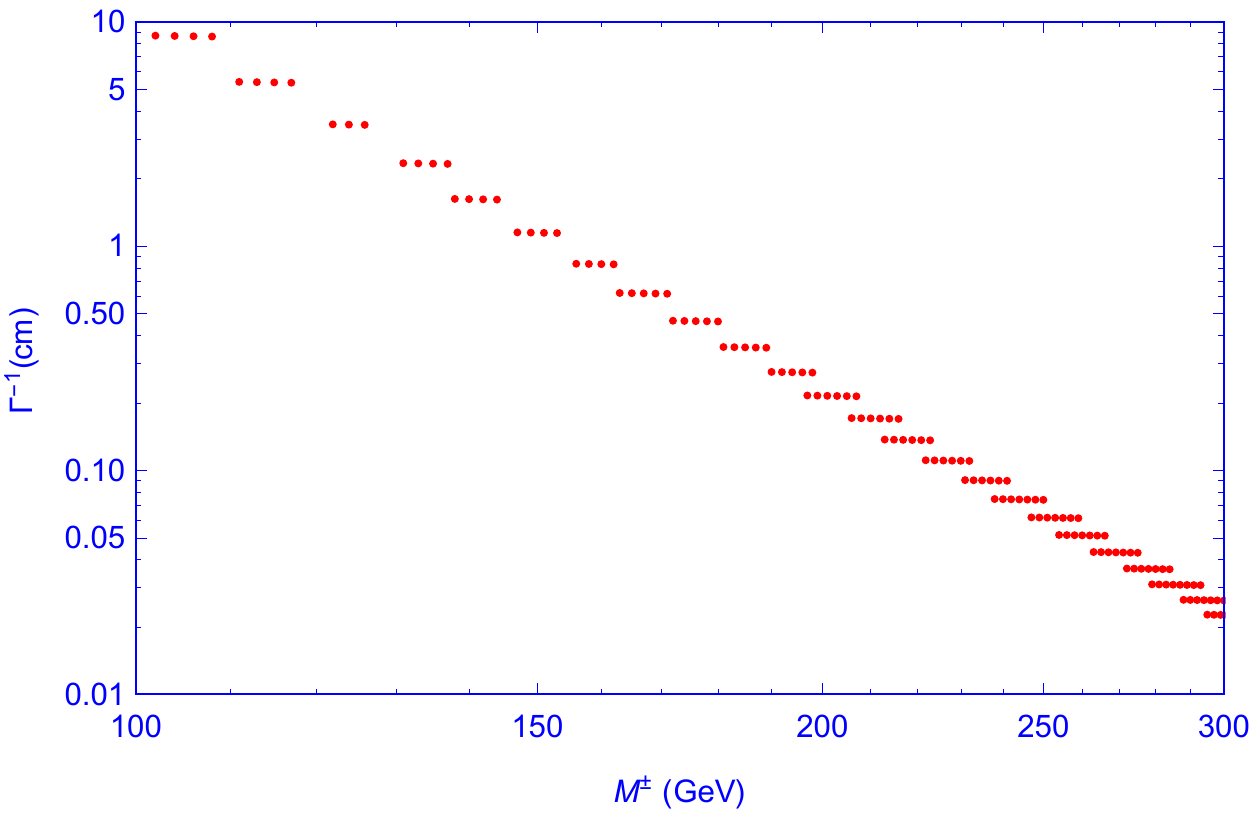}      
$$
	\caption{Left panel: Scatter plot showing relic abundance as
          a function of DM mass with mass splitting less than 50
          GeV. Black solid line shows the correct relic abundance as
          allowed by PLANCK data. Right panel: Displaced vertex ($\Gamma^{-1}$) in cm as a
          function of $M^\pm$ (GeV) for relic density allowed points. Value of mixing angle $\sin \theta
          =3 \times 10^{-4}$ is used in both the plots for illustration.}
\label{fig:DDcon1}
\end{figure}

In the left panel of Fig. \ref{fig:DDcon1}, a scatter plot is shown taking relic abundance  as a
function of DM mass keeping the mass splitting less than $50$ GeV. Here, we fix the singlet-doublet 
mixing angle to be $\sin\theta = 3 \times 10^{-4} $, a moderately smaller value. We have also shown the correct relic 
abundance as allowed by the PLANCK data with a horizontal solid black line. We choose those set of points 
from the relic abundance data which are allowed by the PLANCK result and use them to calculate
the displaced vertex signature of $N^\pm$ ($\Gamma^{-1}$) and plotted as a function of $M^{\pm}$ 
in the right-panel of Fig. (\ref{fig:DDcon1}). We observe that the 
displaced vertex becomes very small for larger values of $M^\pm$, as the inverse of decay width $\Gamma^{-1}$ is 
inversely proportional to the mass of decaying charged particle. However, for smaller masses with $M^\pm \sim 200$ GeV, 
the displaced vertex can be as large as 2.5 mm to be detected in Large Hadron Collider (LHC). The production cross sections 
for such excitations have already been discussed earlier \cite{Bhattacharya:2015qpa} with possible leptonic signatures and 
we refrain from discussing those here again. The important point to be noted here is that to get a large displaced vertex we 
need a small mixing angle between the singlet and doublet. In fact, the small mixing angle is favoured by all the constraints 
we discussed in previous sections, such as correct relic abundance and null detection of DM at direct search experiments. However, 
from Eq. (\ref{theta_constraint}) we also learnt that the singlet-doublet 
mixing can not be arbitrarily small and therefore, the displaced vertex can not be too large.         

\section{Summary and conclusion}\label{con}
We explored the possibility of a singlet-doublet mixed vector-like fermion dark matter in presence
of a scalar triplet. The mixing angle: $\sin\theta$ between the singlet and doublet plays an 
important role in the calculation of relic abundance as well as direct detection. We found that 
the constraint from null detection of DM at direct search experiments and relic abundance can be 
satisfied in a large region of parameter space for mixing angle: $\sin\theta \sim 0.3$ and smaller values. If 
the scalar triplet is light, say $M_\Delta \lesssim 500$ GeV, then it contributes to relic abundance 
only near the resonance i.e with $M_{N_1} \sim \frac{M_\Delta}{2}$. 
On the other hand, if $M_\Delta \gtrsim 1$ TeV, then it decouples and hence 
does not contribute to relic abundance of DM. 

The scalar triplet couples symmetrically to lepton doublets as well as to the doublet component of the DM. 
Therefore, when the scalar triplet acquires an induced vev, it not only gives Majorana masses to the light 
neutrinos but also induce a sub-GeV Majorana mass to the DM. As a result the DM, which was originally a 
vector-like Dirac fermion splits into two pseudo-Dirac fermions with a mass separation of sub-GeV order. 
Due to this reason the Z-mediated inelastic scattering of the DM with nucleon is suppressed. However, we found that 
the spin independent direct detection of DM through the SM Higgs mediation is in the right ballpark of 
Xenon-1T. The absence of $Z$ mediated DM-nucleon cross-section relaxes the constraint on mixing angle 
$\sin \theta$ as we can go as high as $\sin \theta =0.3$ for DM mass
$M_1 > 400$ GeV for small mass splitting $\Delta M < 100$ GeV.  This high value of $\sin \theta$ is also well satisfied
  by the correct relic abundance. So the spin independent direct detection cross-section does not put stronger constraint 
on the mixing angle if the mass splitting is not so large and allows large region of parameter space 
unlike the model in absence of a triplet.

The $\rho$ parameter in the SM restricts the vev of scalar triplet to $u_\Delta \leq 3.64$ GeV. 
This in turn gives the mixing between the SM Higgs and $\Delta$ to be $\sin\theta_0$ $\mathcal{O}(10^{-2})$ 
even if the $M_\Delta \lesssim 500$ GeV. Therefore, $\Delta$ does not contribute significantly to 
the spin independent direct detection cross-section. 
\\
\begin{figure}[thb]
$$
	\includegraphics[height=6.0cm]{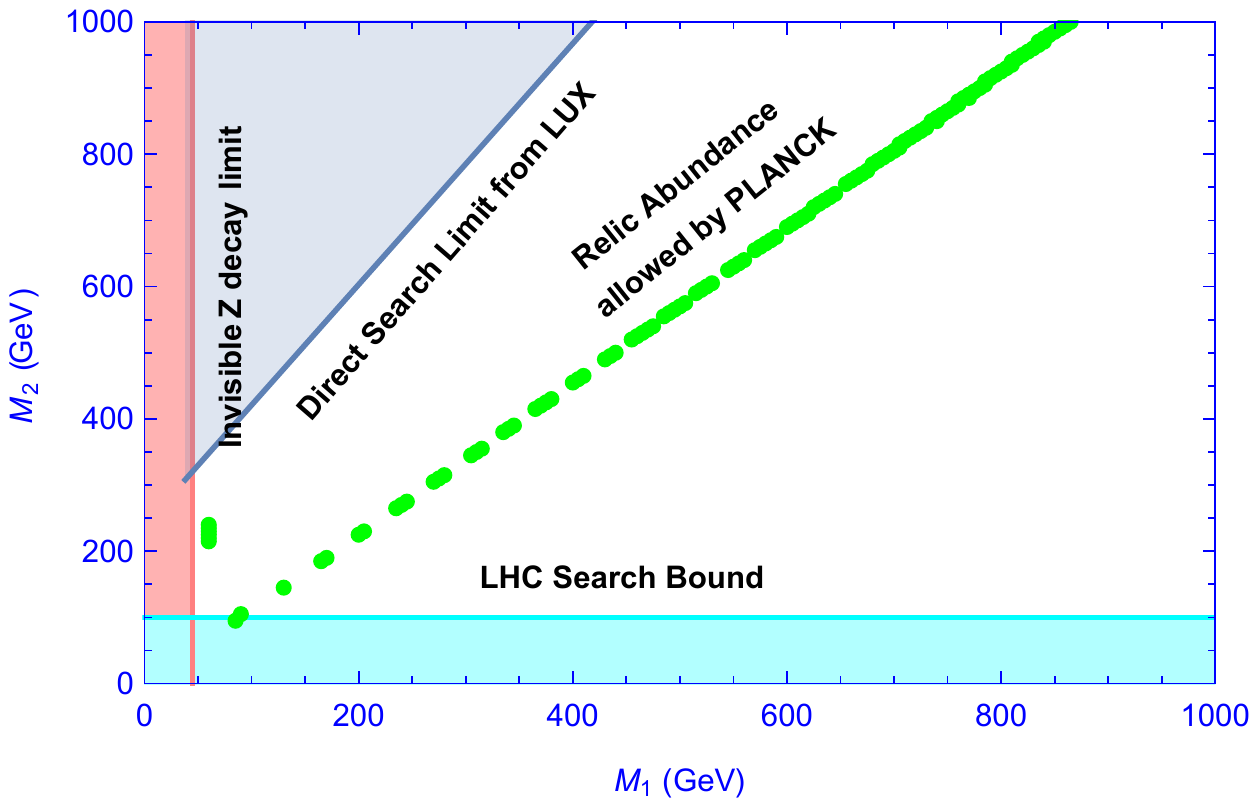}
	\includegraphics[height=6.0cm]{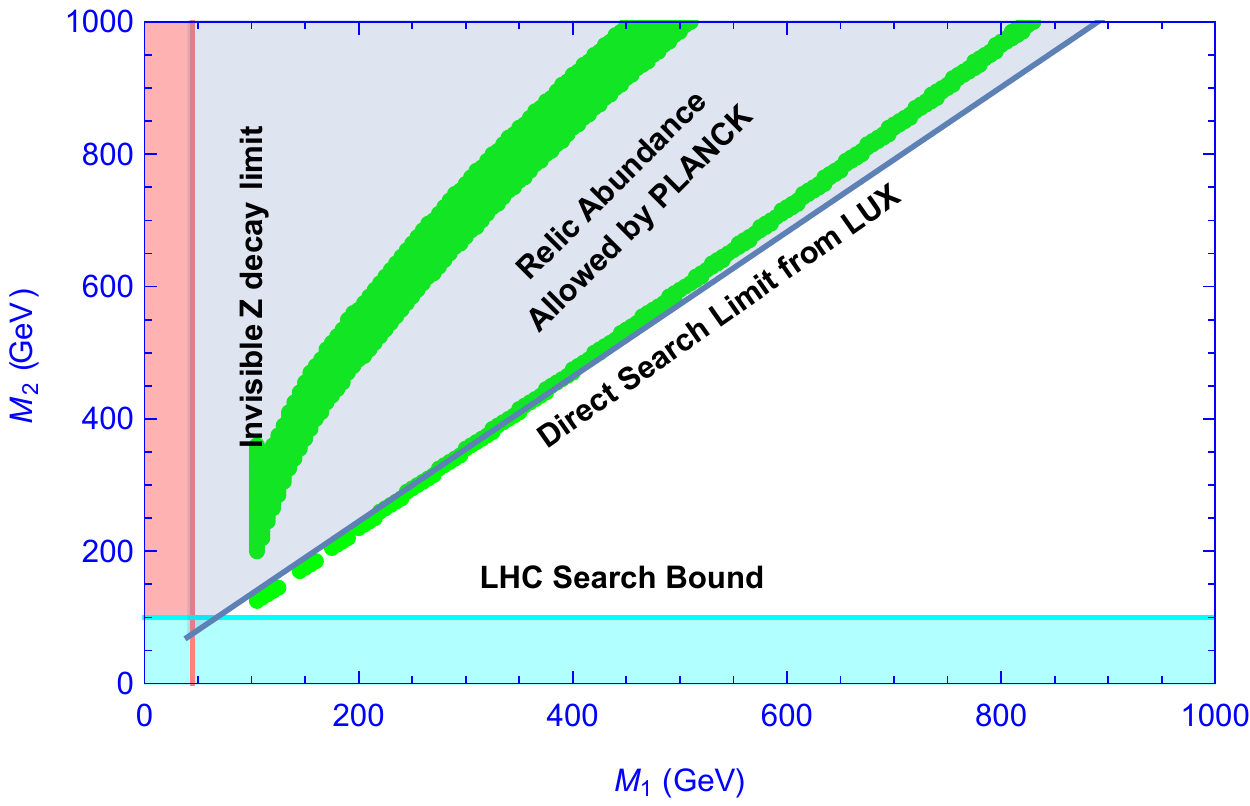}
$$
	\caption{Summary of all constraints in the plane of $M_1 -
          M_2$ using $\sin \theta=0.1$ (left) and $\sin\theta=0.3$ (right). }
\label{fig:summary}
\end{figure}
\\
We summarize the constraints on the parameters in Fig.~\ref{fig:summary}, where we have 
shown the allowed values in the plane of $M_1-M_2$ using $\sin\theta=0.1$ in the left and for $\sin\theta=0.3$ on the right. 
The green points are allowed by the relic abundance of DM by taking the constraint from PLANCK data. The main constraint comes from 
non observation of DM from direct search data of LUX experiment. On the left, for small $\sin\theta=0.1$, direct search constraint is 
less severe as has already been discussed and the whole relic density allowed points are consistent with direct search bound. However, 
for larger $\sin\theta=0.3$, on the right hand side of Fig.~\ref{fig:summary}, a significant part of the relic density allowed space is submerged 
into direct search bound excepting for the low DM mass region upto $\sim$ 400 GeV. The direct search bound gets more 
stringent with larger $\Delta M$ and that is one of the primary reasons that relic density allowed parameter space 
with large $\sin\theta=0.3$ is disfavoured. This is still significantly new in contrast to the model without the 
triplet, where larger mixing $\ge 0.1$, was completely forbidden by direct search data.  
There are other small regions which are disfavoured by various experimental searches. For example, the region in cyan 
colour is disfavoured by the collider search of $N^\pm$ and hence the allowed values are given by 
$M^\pm \sim M_2 > 100 \rm GeV$. The mass of $N_1$ (DM), i.e., $M_1 > 45$ GeV, is required in order 
to relax the severe constraints from the invisible $Z$ boson decay \cite {Bhattacharya:2015qpa}. 
The charged partner of the DM gives interesting signatures at colliders if $M^\pm  - M_1 \lesssim 80$ GeV. 
As a result the two body decay of $N^\pm$ is forbidden. The only way it can decay is the three body decay. For example, 
the notable one is $N^- \to N_1 \ell^- \overline{\nu_\ell}$. In the small singlet-doublet mixing limit we get a
displaced vertex of $10$ cm for $M^\pm  \sim 100$ GeV and a mass splitting of few tens of GeV while satisfying the constraint 
from observed relic abundance.

\section* { Acknowledgement}
 SB would like to acknowledge the DST-INSPIRE project with grant no. IFA-13 PH-57  at IIT Guwahati.


\begin{thebibliography}{99}
\bibitem{DM_review1} 
G. Bertone, D. Hooper and J. Silk, 
{\it Particle Dark Matter: Evidence, Candidates and Constraints}, 
Phys. Rept. 405, 279 (2005), arXiv:hep-ph/0404175.

\bibitem{DM_review2}
G. Jungman, M. Kamionkowski and K. Griest, 
{\it Supersymmetric Dark Matter}, 
Phys. Rept. 267, 195 (1996), arXiv:hep-ph/9506380.

\bibitem{wmap} 
G.~Hinshaw {\it et al.} [WMAP Collaboration],
  Astrophys.\ J.\ Suppl.\  {\bf 208}, 19 (2013)
  [arXiv:1212.5226 [astro-ph.CO]].
\bibitem{PLANCK} 
P.~A.~R.~Ade {\it et al.}  [Planck Collaboration],
Astron.\ Astrophys.\  {\bf 571}, A16 (2014), arXiv:1303.5076 [astro-ph.CO].
\bibitem{Fukuda:2001nk} 
  S.~Fukuda {\it et al.} [Super-Kamiokande Collaboration],
  Phys.\ Rev.\ Lett.\  {\bf 86}, 5656 (2001)
  [hep-ex/0103033].


\bibitem{Bilenky:1980cx} 
  S.~M.~Bilenky, J.~Hosek and S.~T.~Petcov,
  Phys.\ Lett.\  {\bf 94B}, 495 (1980).


\bibitem{type_I} P. Minkowski, Phys. Lett. B 67 (1977) 421; M. Gell-Mann, P. Ramond and R. Slansky, Pro-
ceedings of the Supergravity Stony Brook Workshop, eds. P. van Niewenhuizen and D. Freed-
man (New York, 1979); T. Yanagida, Proceedings of the Workshop on the Baryon Number of
the Universe and Unified Theories, Tsukuba, Japan, 13-14 Feb 1979; R. N. Mohapatra and
G. Senjanovic, Phys. Rev. Lett. 44, 912 (1980) .

\bibitem{type_II}
  M.~Magg and C.~Wetterich,
  Phys.\ Lett.\  {\bf 94B}, 61 (1980).
  G.~Lazarides, Q.~Shafi and C.~Wetterich,
  Nucl.\ Phys.\ B {\bf 181}, 287 (1981).
  R.~N.~Mohapatra and G.~Senjanovic,
  Phys.\ Rev.\ D {\bf 23}, 165 (1981).
  E.~Ma and U.~Sarkar,
  Phys.\ Rev.\ Lett.\  {\bf 80}, 5716 (1998)
  [hep-ph/9802445];
  W.~Konetschny and W.~Kummer,
  Phys.\ Lett.\  {\bf 70B}, 433 (1977).
  J.~Schechter and J.~W.~F.~Valle,
  Phys.\ Rev.\ D {\bf 22}, 2227 (1980).
  T.~P.~Cheng and L.~F.~Li,
  Phys.\ Rev.\ D {\bf 22}, 2860 (1980).

\bibitem{type_III}R. Foot, H. Lew, X. G. He and G. C. Joshi, Z. Phys. C 44, 441 (1989)


\bibitem{dim-5-operator} S. Weinberg, Phys. Rev. Lett. 43, 1566, 1979


\bibitem{Bhattacharya:2016rqj} 
  S.~Bhattacharya, B.~Karmakar, N.~Sahu and A.~Sil,
  arXiv:1611.07419 [hep-ph].

\bibitem{Bhattacharya:2016lts} 
  S.~Bhattacharya, B.~Karmakar, N.~Sahu and A.~Sil,
  Phys.\ Rev.\ D {\bf 93}, no. 11, 115041 (2016)
  [arXiv:1603.04776 [hep-ph]].

\bibitem{Bhattacharya:2016qsg} 
  S.~Bhattacharya, S.~Jana and S.~Nandi,
  Phys.\ Rev.\ D {\bf 95}, no. 5, 055003 (2017)
  [arXiv:1609.03274 [hep-ph]].

\bibitem{Sahu:2007uh} 
  N.~Sahu and U.~Sarkar,
  Phys.\ Rev.\ D {\bf 76}, 045014 (2007)
  [hep-ph/0701062].

\bibitem{McDonald:2007ka} 
  J.~McDonald, N.~Sahu and U.~Sarkar,
  JCAP {\bf 0804}, 037 (2008)
[arXiv:0711.4820 [hep-ph]].

\bibitem{Sahu:2008aw} 
  N.~Sahu and U.~Sarkar,
  Phys.\ Rev.\ D {\bf 78}, 115013 (2008)
  doi:10.1103/PhysRevD.78.115013
  [arXiv:0804.2072 [hep-ph]].

\bibitem{Chatterjee:2014vua} 
  A.~Chatterjee and N.~Sahu,
  Phys.\ Rev.\ D {\bf 90}, no. 9, 095021 (2014)
  doi:10.1103/PhysRevD.90.095021
  [arXiv:1407.3030 [hep-ph]].

\bibitem{Patra:2014sua} 
  S.~Patra, N.~Sahoo and N.~Sahu,
  Phys.\ Rev.\ D {\bf 91}, no. 11, 115013 (2015)
  doi:10.1103/PhysRevD.91.115013
  [arXiv:1412.4253 [hep-ph]].

\bibitem{Patra:2016shz} 
  S.~Patra, S.~Rao, N.~Sahoo and N.~Sahu,
  Nucl.\ Phys.\ B {\bf 917}, 317 (2017)
  doi:10.1016/j.nuclphysb.2017.02.010
  [arXiv:1607.04046 [hep-ph]].


\bibitem{Bhattacharya:2015qpa} 
  S.~Bhattacharya, N.~Sahoo and N.~Sahu,
  Phys.\ Rev.\ D {\bf 93}, no. 11, 115040 (2016)
  [arXiv:1510.02760 [hep-ph]].

\bibitem{Bhattacharya:2016lyg} 
  S.~Bhattacharya, S.~Patra, N.~Sahoo and N.~Sahu,
  JCAP {\bf 1606}, no. 06, 010 (2016)
  [arXiv:1601.01569 [hep-ph]].






\bibitem{ArkaniHamed:2005yv} 
  N.~Arkani-Hamed, S.~Dimopoulos and S.~Kachru,
  hep-th/0501082.

\bibitem{Mahbubani:2005pt} 
  R.~Mahbubani and L.~Senatore,
  Phys.\ Rev.\ D {\bf 73}, 043510 (2006)
  [hep-ph/0510064].

\bibitem{D'Eramo:2007ga} 
  F.~D'Eramo,
  Phys.\ Rev.\ D {\bf 76}, 083522 (2007)
  [arXiv:0705.4493 [hep-ph]].

\bibitem{Enberg:2007rp} 
  R.~Enberg, P.~J.~Fox, L.~J.~Hall, A.~Y.~Papaioannou and M.~Papucci,
  JHEP {\bf 0711}, 014 (2007)
  [arXiv:0706.0918 [hep-ph]].

\bibitem{Cynolter:2008ea} 
  G.~Cynolter and E.~Lendvai,
  Eur.\ Phys.\ J.\ C {\bf 58}, 463 (2008)
  [arXiv:0804.4080 [hep-ph]].

\bibitem{Cohen:2011ec} 
  T.~Cohen, J.~Kearney, A.~Pierce and D.~Tucker-Smith,
  Phys.\ Rev.\ D {\bf 85}, 075003 (2012)
  [arXiv:1109.2604 [hep-ph]].

\bibitem{Cheung:2013dua} 
  C.~Cheung and D.~Sanford,
  JCAP {\bf 1402}, 011 (2014)
  [arXiv:1311.5896 [hep-ph]].

\bibitem{Restrepo:2015ura} 
  D.~Restrepo, A.~Rivera, M.~Sánchez-Peláez, O.~Zapata and W.~Tangarife,
  Phys.\ Rev.\ D {\bf 92}, no. 1, 013005 (2015)
  [arXiv:1504.07892 [hep-ph]].

\bibitem{Calibbi:2015nha} 
  L.~Calibbi, A.~Mariotti and P.~Tziveloglou,
  arXiv:1505.03867 [hep-ph].

\bibitem{Cynolter:2015sua} 
  G.~Cynolter, J.~Kovacs and E.~Lendvai,
  arXiv:1509.05323 [hep-ph].

\bibitem{xenon100}
  E.~Aprile {\it et al.} {\bf XENON100 Collaboration},
  ``Dark Matter Results from 225 Live Days of XENON100 Data,''
  Phys.\ Rev.\ Lett.\  {\bf 109}, 181301 (2012)
  [arXiv:1207.5988 [astro-ph.CO]].

\bibitem{Akerib:2016vxi} 
  D.~S.~Akerib {\it et al.},
  arXiv:1608.07648 [astro-ph.CO].

\bibitem{Aprile:2015uzo} 
  E.~Aprile {\it et al.} [XENON Collaboration],
  JCAP {\bf 1604}, no. 04, 027 (2016)
  [arXiv:1512.07501 [physics.ins-det]].




\bibitem{pdg} 
 C. Patrignani et al. (Particle Data Group), Chin. Phys. C40, 100001
 (2016).
\bibitem{Cheung:2015dta} 
  K.~Cheung, P.~Ko, J.~S.~Lee and P.~Y.~Tseng,
  JHEP {\bf 1510}, 057 (2015)
  [arXiv:1507.06158 [hep-ph]].'
\bibitem{Hartling:2014aga} 
  K.~Hartling, K.~Kumar and H.~E.~Logan,
  Phys.\ Rev.\ D {\bf 91}, no. 1, 015013 (2015)
  [arXiv:1410.5538 [hep-ph]].
\bibitem{Keung:1984hn} 
  W.~Y.~Keung and W.~J.~Marciano,
  Phys.\ Rev.\ D {\bf 30}, 248 (1984).

\bibitem{Dilepton}See for a partial list: 
 A.~G.~Akeroyd, M.~Aoki and H.~Sugiyama,
  Phys.\ Rev.\ D {\bf 77}, 075010 (2008)
  [arXiv:0712.4019 [hep-ph]],
J.~Garayoa and T.~Schwetz,
  JHEP {\bf 0803}, 009 (2008)
  [arXiv:0712.1453 [hep-ph]],
E.~J.~Chun, K.~Y.~Lee and S.~C.~Park,
  Phys.\ Lett.\ B {\bf 566}, 142 (2003)
  [hep-ph/0304069],
M.~Kadastik, M.~Raidal and L.~Rebane,
  Phys.\ Rev.\ D {\bf 77}, 115023 (2008)
  [arXiv:0712.3912 [hep-ph]],
P.~Fileviez Perez, T.~Han, G.~y.~Huang, T.~Li and K.~Wang,
  Phys.\ Rev.\ D {\bf 78}, 015018 (2008)
  [arXiv:0805.3536 [hep-ph]],
S.~K.~Majee and N.~Sahu,
  Phys.\ Rev.\ D {\bf 82}, 053007 (2010)
  [arXiv:1004.0841 [hep-ph]].

\bibitem{Arina:2011cu} 
  C.~Arina and N.~Sahu,
  Nucl.\ Phys.\ B {\bf 854}, 666 (2012)
  [arXiv:1108.3967 [hep-ph]].
\bibitem{Arina:2012fb} 
  C.~Arina, J.~O.~Gong and N.~Sahu,
  Nucl.\ Phys.\ B {\bf 865}, 430 (2012)
  [arXiv:1206.0009 [hep-ph]].

\bibitem{Arina:2012aj} 
  C.~Arina, R.~N.~Mohapatra and N.~Sahu,
  Phys.\ Lett.\ B {\bf 720}, 130 (2013)
  [arXiv:1211.0435 [hep-ph]].
\bibitem{micro} G.~Belanger, F.~Boudjema, A.~Pukhov and A.~Semenov,
  Comput.\ Phys.\ Commun.\  {\bf 180}, 747 (2009)
  [arXiv:0803.2360 [hep-ph]].
\bibitem{griest}
K.~Griest and D.~Seckel, {\it Three exceptions in the calculation of relic abundances}, 
Phys.\ Rev.\ {\bf D\,43}, 3191 (1991);

\bibitem{TuckerSmith:2001hy} 
  D.~Tucker-Smith and N.~Weiner,
  Phys.\ Rev.\ D {\bf 64}, 043502 (2001)
  [hep-ph/0101138].
\bibitem{Cui:2009xq} 
  Y.~Cui, D.~E.~Morrissey, D.~Poland and L.~Randall,
  JHEP {\bf 0905}, 076 (2009)
  [arXiv:0901.0557 [hep-ph]].
















\bibitem{Goodman:1984dc} 
  M.~W.~Goodman and E.~Witten,
  Phys.\ Rev.\ D {\bf 31}, 3059 (1985).
\bibitem{Essig:2007az} 
  R.~Essig,
  Phys.\ Rev.\ D {\bf 78}, 015004 (2008)
  [arXiv:0710.1668 [hep-ph]].







\end{thebibliography}
\end{document}